\documentclass[preprint,final]{aastex}
\shorttitle{Variables in "Kepler Field"}
\shortauthors{Hartman et al.}
\begin{document}

\title{HAT Variability Survey in the High Stellar Density ``Kepler
Field'' with Millimagnitude Image Subtraction Photometry.}

\author{J.~D.~Hartman, G.~Bakos\altaffilmark{1,2}, K.~Z.~Stanek, R.~W.~Noyes}
\affil{Harvard-Smithsonian Center for Astrophysics, 60 Garden St.,
Cambridge, MA~02138}
\email{jhartman, gbakos, kstanek, rnoyes@cfa.harvard.edu}

\altaffiltext{1}{Predoctoral Fellow, Smithsonian Astrophysical Observatory}
\altaffiltext{2}{Also at Konkoly Observatory, Budapest, H-1525, P.O.~Box 67}

\begin{abstract}

The Hungarian-made Automated Telescope network (HATnet) is an ongoing
project to detect transiting extra-solar planets using small aperture
(11 cm diameter), robotic telescopes. In this paper we present the
results from using image subtraction photometry to reduce a crowded
stellar field observed with one of the HATnet telescopes (HAT-5). This
field was chosen to overlap with the planned {\it Kepler}\/
mission. We obtained I-band light curves for 98,000 objects in a
67-square-degree field of view centered at $(\alpha,\delta)=(19^{\rm
h}44^{\rm m}00\fs0, +37\arcdeg32\arcmin00\farcs0)$ (J2000.0), near the
Galactic plane in the constellations Cygnus and Lyra. These
observations include 788 5-minute exposures over 30 days. For the
brightest stars (I $\sim$ 8.0) we achieved a precision of 3.5
millimagnitudes, falling to 0.1 magnitudes at the faint end (I $\sim$
14). From these light curves we identify 1617 variable stars, of which
1439 are newly discovered. The fact that nearly $90$\% of the
variables were previously undetected further demonstrates the vast
number of variables yet to be discovered even among fairly bright
stars in our Galaxy. We also discuss some of the most interesting
cases. This includes: V1171 Cyg, a triple system with the inner two
stars in a $P=1.462\;$day period eclipsing orbit and the outer star a
$P=4.86\;$day Cepheid; HD227269, an eccentric eclipsing system with a
$P=4.86\;$day period that also shows $P=2.907\;$day pulsations; WW
Cyg, a well studied eclipsing binary; V482 Cyg, an RCB star; and V546
Cyg, a PV Tel variable. We also detect a number of small amplitude
variables, in some cases with full amplitude as low as 10 mmag.

\end{abstract} 

\section{Introduction}

The Hungarian-made Automated Telescope network (HATnet) is an ongoing
project to detect transiting extra-solar planets using small aperture
(11 cm diameter), robotic telescopes (Bakos et al.\ 2004, hereafter
B04). The HATnet telescopes make use of a fast focal ratio (f/1.8) to
efficiently monitor a large number of fairly bright stars ($I<14.5$)
over a wide field-of-view (FOV). A number of other groups have also
taken this small-telescope approach toward finding transits (see
Horne\ 2003 for a comprehensive list). In contrast, there are several
groups that employ a ``narrow, but deep'' method. This includes the
transit search by the Optical Gravitational Lensing Experiment (OGLE)
project, which to date has discovered three confirmed ``very hot
Jupiters'' (Udalski et al.\ 2002, 2003; Konacki et al.\ 2003; Torres
et al.\ 2004; Bouchy et al.\ 2004; Konacki et al.\ 2004), the only
planets detected so far by transit searches.

Besides the size of the telescopes and FOV, another difference between
the approaches is the method used to obtain photometry for the
monitored stars. For several years many of the ``narrow and deep''
searches have made use of the image subtraction techniques due to
Alard \& Lupton\ (1998; also Alard\ 2000). This includes OGLE which
uses Difference Image Analysis (DIA, Wozniak\ 2000), and PISCES
(Mochejska et al.\ 2002, 2004) which uses a different implementation
of image subtraction in monitoring the open clusters NGC~2158 \&
NGC~6791.

Image subtraction is the current state of the art for massive
time-series photometry. It has been shown that in narrow, dense fields,
it can produce light curves with precision down to the photon limit
(e.g.~see Mochejska et al.\ 2002). However, to date there exists no
published results that use image subtraction in a wide-field setting.
This has limited these searches to observing only relatively isolated
stars in regions where point spread function (PSF) fitting and aperture
photometry yield high precision.

In this paper we report our use of image subtraction to obtain light
curves for 98,000 objects in a single field, near the galactic plane,
observed with one of the HATnet telescopes (HAT-5).  This field was
chosen in particular because of its overlap with NASA's {\it Kepler}\/
mission to observe transiting planets from space (Borucki et al.\
2003). Over 9,000 of the brightest light curves have a
root-mean-square (RMS) of less than 1\% (i.e. better than 10
millimagnitude precision) at 5-min sampling.  In the following section
we describe our observations, and in \S3 we discuss our image
subtraction based data reduction to obtain the light curves.

While the main purpose of HATnet remains the discovery of transiting
extra-solar planets, it is also useful for discovering and
characterizing variable stars in the Galaxy. To this end we have
analyzed these light curves to select a list of 1617 variable stars, of
which 1439 are newly discovered. We describe our selection criteria in
\S4 and present our catalog, including a discussion of many interesting
cases, in \S5. We finish with a brief summary of our results in \S6.

\section{Observations}

The data were obtained in June and July, 2003 using the HAT-5 telescope
located at the Fred Lawrence Whipple Observatory (FLWO). The telescope
uses a Canon 11~cm diameter f/1.8L lens and a Cousins I-band filter to image
onto an Apogee AP10 front-illuminated, 2K$\times$2K CCD. The result is
an $8.3^{\circ}\times 8.3^{\circ}$~FOV image with a pixel scale of
$14\arcsec$. For details on the design and performance of the
instrument see B04. The pointing was stepped in a
prescribed pattern of sub-pixel increments during each exposure to
broaden the full width at half maximum (FWHM) of the stellar profiles
from $\sim 1.5$~pix to $\sim 2.5$~pix.

\begin{figure}[p]
\epsscale{1}
\plotone{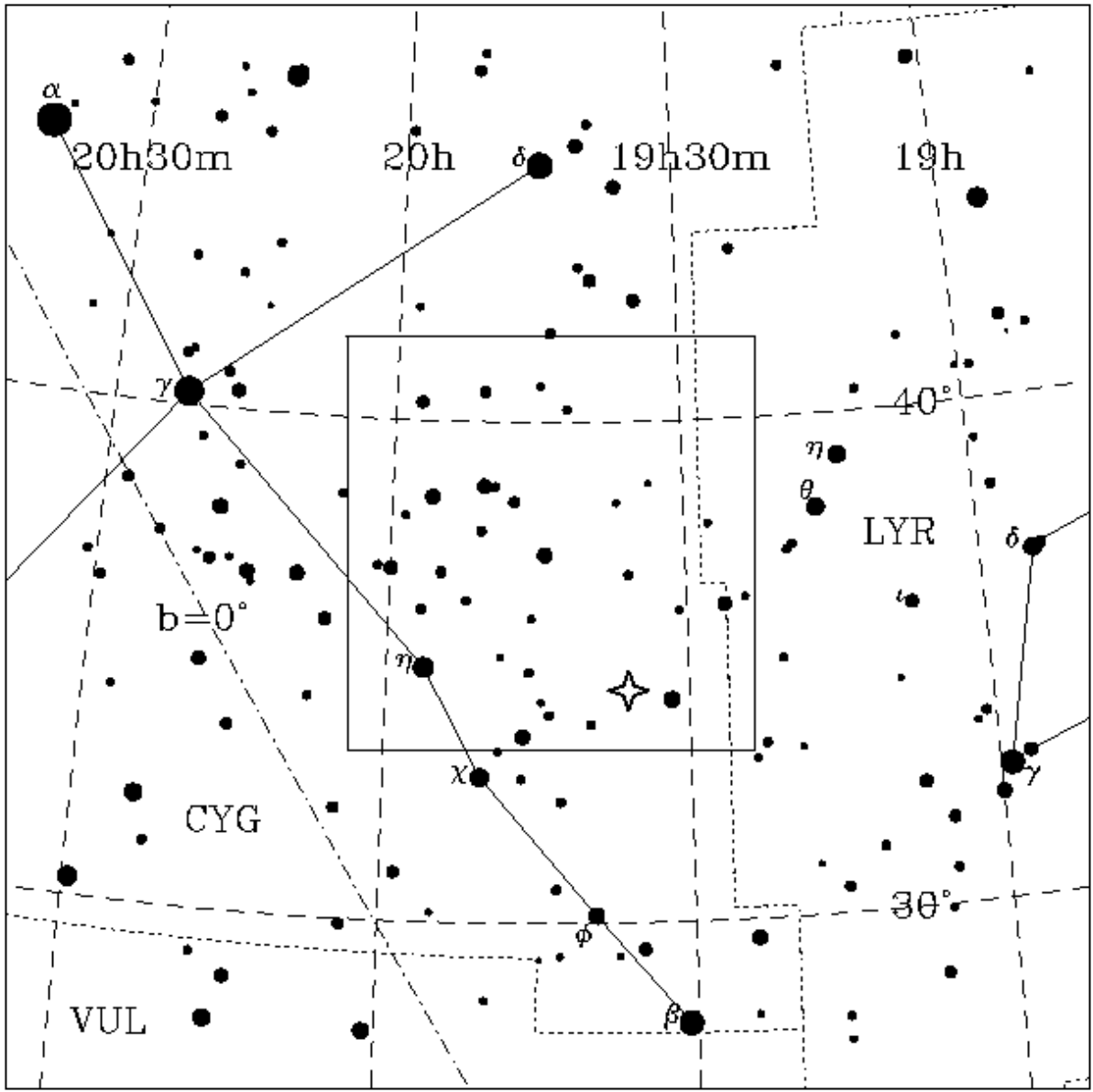}
\caption{The observed $8.3\arcdeg$x$8.3\arcdeg$ field of view, lying at
the western boundary of Cygnus. The proposed center of the {\it
Kepler}\/ Mission is marked with a star. The plane of the Galactic disk
is shown with the dot-dashed line.}
\label{fov}
\end{figure}
\begin{figure}[p]
\epsscale{1}
\plotone{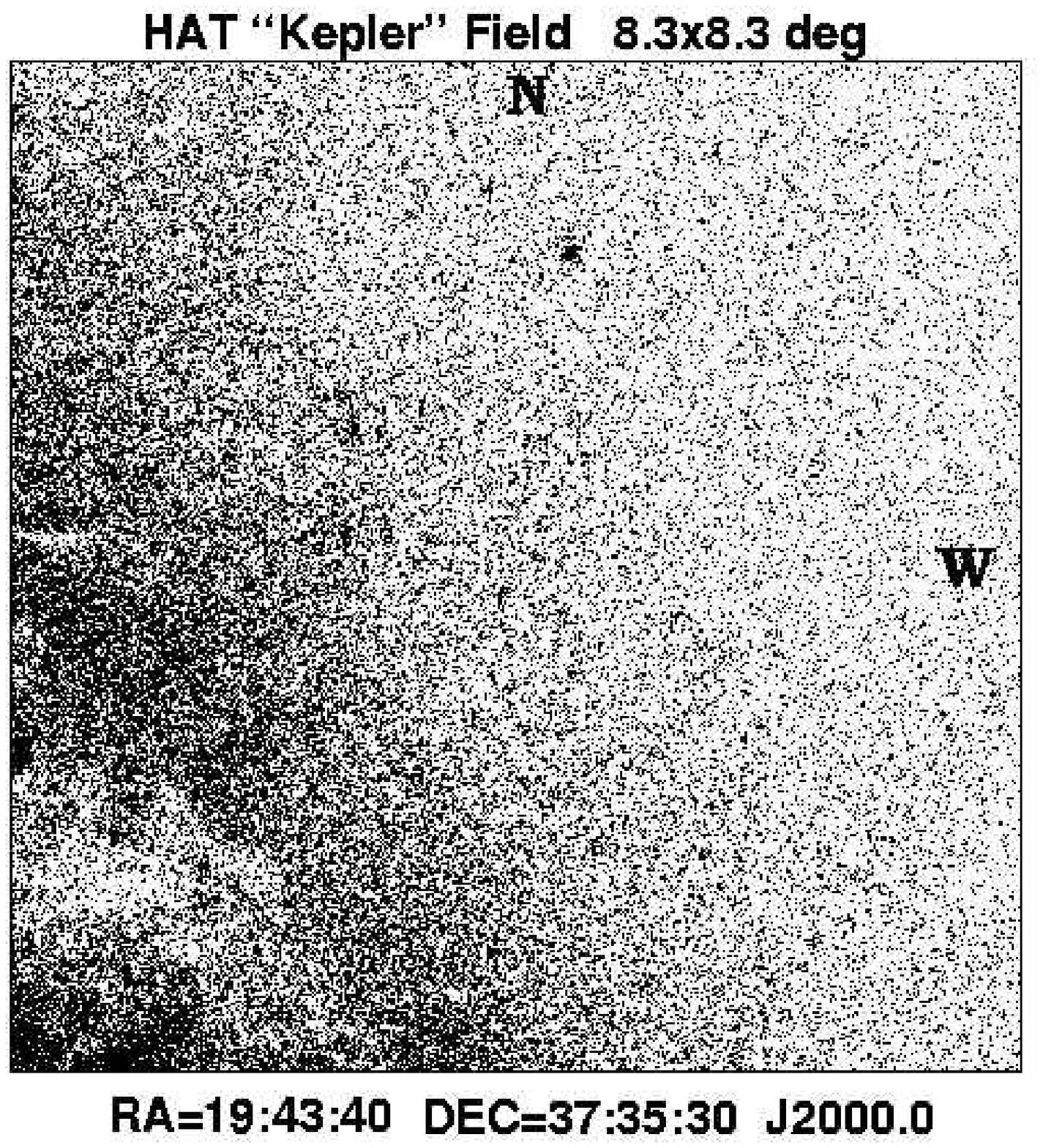}
\caption{Image of the HAT ``Kepler'' field (HAT-199). Note the very
high stellar density, and variable extinction, particularly toward the
Galactic disk in the lower left corner.}
\label{refimage}
\end{figure} 

The field we observed (HAT-199) is centered at 
$(\alpha,\delta)=(19^{\rm h}44^{\rm m}00\fs0, +37\arcdeg32\arcmin00\farcs0)$ 
(J2000.0) and lies at the western boundary of the constellation Cygnus
(Fig.~\ref{fov}). The south-eastern (lower left) corner of the field is
within $2^{\circ}$ of the Galactic plane. As mentioned above, this
field was chosen to overlap with the proposed center for NASA's {\it
Kepler}\/ Mission (D.~Latham, private communication). Thus these
observations may be useful as a means of identifying interesting
objects (including stars bearing transiting planets) to be investigated
at higher precision with {\it Kepler}.

This field contains hundreds of thousands of relatively bright
sources, from which we select 98,000 with $I<14.8.$ Just from the
sources we select the density is 0.4 objects per square-arcmin. At our
pixel scale this corresponds to one object per 42 pixels, or a typical
separation of 6-7 pixels between the objects. At this density the
stellar profiles are highly blended, particularly toward the galactic
plane (Fig.~\ref{refimage}). Because of this severe blending, large
portions of this field would not be useful for transit searches or
precision photometry in general using traditional photometric methods
(e.g. simple aperture photometry, or PSF fitting).

\section{Data Reduction}

\subsection{Image Subtraction}

The preliminary CCD reductions including dark current subtraction,
flat-fielding, etc.~were discussed in B04. 

To obtain photometry we used the image subtraction methods due to
Alard \& Lupton\ (1998, also Alard 2000). We describe the procedure
here, referring the reader interested in the theoretical basis of the
method to the original papers.

The simplest method to measure the apparent magnitude of a star is to
measure the total signal within a fixed aperture centered on the star
using a constant weight per pixel. However, when one is working on a
crowded stellar field this method breaks down since more than one
source will contribute light to the aperture. The typical procedure in
this case is to determine a PSF for the frame using bright/isolated
stars, and then fit that PSF to all the stars on the frame.

When obtaining light curves for the stars in a dense field, it is
useful to note that on average the intrinsic magnitude of the stars
does not change from image to image. Instead of performing PSF fitting
separately on every image, one can obtain higher precision by
performing weighted aperture photometry on the sparsely populated
difference image formed by subtracting the image from a reference
image. To do this one needs a method for matching two images with
different PSFs before subtracting so that the subtracted image is not
dominated by residuals from the different PSFs. Alard \& Lupton (1998)
solved this problem by proposing an efficient method for finding a
function that transforms the PSF of the reference image into the PSF
of the image to be subtracted (the kernel). Their scheme does not
assume anything about the shape of the PSF, only that the kernel,
which relates the PSFs on different images, can be written in terms of
a product of gaussians and polynomials (of arbitrary order). Alard
(2000) modified the procedure to allow for spatial variations in the
kernel which can be fit with a polynomial. When fitting the kernel it
is not necessary to use only bright/isolated stars, instead even stars
in the highest density regions contribute to the fit. This is one
reason why the kernel can be obtained with better accuracy than an
independent determination of the PSF. Another reason why the kernel
can be determined with better accuracy than the PSF is that while the
PSF can have an arbitrary shape that may not be well-fit by any model
function, empirically the kernel relating two PSFs from the same
instrument appears to be well-fit by Alard \& Lupton's model,
regardless of the shape of the PSF. Moreover, because the method
allows for a constant scaling of the kernel, any correlated variations
such as the change in magnitude due to different atmospheric
extinction are automatically removed.

Once the kernel is obtained and the images have been subtracted, it is
a simple step to obtain the relative change in magnitude between the
subtracted image and the reference image. This is done by determining
a PSF for the reference image, convolving it with the kernel, and then
performing weighted aperture photometry on the subtracted image. Since
the kernel can be obtained with better accuracy than the PSF, by
determining a single PSF and applying it to all the images (after
transforming with the kernel) one can achieve higher precision light
curves than by determining the PSF of each image separately. Note that
this procedure only works in the regime where the PSF can be
accurately determined on the reference image. In the limit of an
extremely crowded field, one may have to obtain the PSF directly on
the subtracted image.

All of the above procedures are included in the ISIS 2.1
package\footnote{ISIS package is available from C. Alard's website at
http://www2.iap.fr/users/alard/package.html}. There are several
references on how to run ISIS, the procedure we follow is similar to
that used by Mochejska et al.\ (2002). We discuss the specifics of our
implementation in Appendix A.

\subsection{Photometry}

There are two approaches that one can take toward obtaining light
curves and identifying variables with subtracted images. The first is
to co-add the absolute value of the subtracted images and search for
strong point source signals to identify variable stars for which to
obtain light curves.  The second is to generate a list of stars from
the reference image, measure photometry on the subtracted images for a
selection of stars on the list, and scan the resulting light curves
for variables. Using the former approach it becomes very difficult to
efficiently identify extremely subtle variations, such as those due to
a transiting planet, as these tend to get washed out in the overall
noise of the image (Mochejska et al.\ 2002). Since the transit
candidates are typically amongst the sources with the smoothest light
curves, it is more efficient to obtain photometry for all the objects
and use routines that are optimized for selecting candidates directly
from the light curves. For this reason we implement the latter
approach. In doing so we do make two sacrifices: any transient
phenomena which do not have a signal in the 47 images combined to make
the reference image go unnoticed, and constant stars located near
variable stars will have variable light curves as a result of the
non-deblending, weighted aperture photometry performed on the
subtracted images by the ISIS routine ``phot.csh.'' The second problem
turns out to be significant for our program, and we describe the steps
we have taken to mitigate it in \S4.6. Note that this procedure also
requires the ability to generate an input list of stars. Since image
subtraction is often able to perform even in the densest fields, when
working in an extremely crowded field the former approach may be the
only option for identifying variables

To obtain the list of stars we used the DAOPHOT/ALLSTAR package
(Stetson 1987, 1992). The list contains 98,000 objects ranging in
magnitude from $I$=7.79 down to 14.87. We then obtained light curves
for all 98,000 objects using the ``phot.csh'' routine contained in the
ISIS package. We discuss the details of this procedure in Appendix B.

We used the {\it Hipparcos}\/ main catalogue (Perryman et al.\ 1997)
to provide the absolute calibration for our instrumental, I-band,
reference magnitudes. Using the coordinates we obtained from matching
to the Two Micron All Sky Survey ({\it 2MASS}\/; Skrutskie et al.\
1997) as discussed in \S3.4, we obtained matches with 55 point sources
from {\it Hipparcos}\/ that also had I-band measurements listed. The I
magnitudes listed were all obtained from ground-based measurements, or
from transformations of {\it Hipparcos}\/ measurements in other
photometric systems. We find a 1-$\sigma$ uncertainty of $\pm 0.06$
magnitudes on our absolute calibration to {\it Hipparcos}. This error
is likely on the {\it Hipparcos} side and may be due to the variety of
sources for the I-band measurements.

\subsection{Photometric Precision}

\begin{figure*}[p]
\epsscale{1}
\plotone{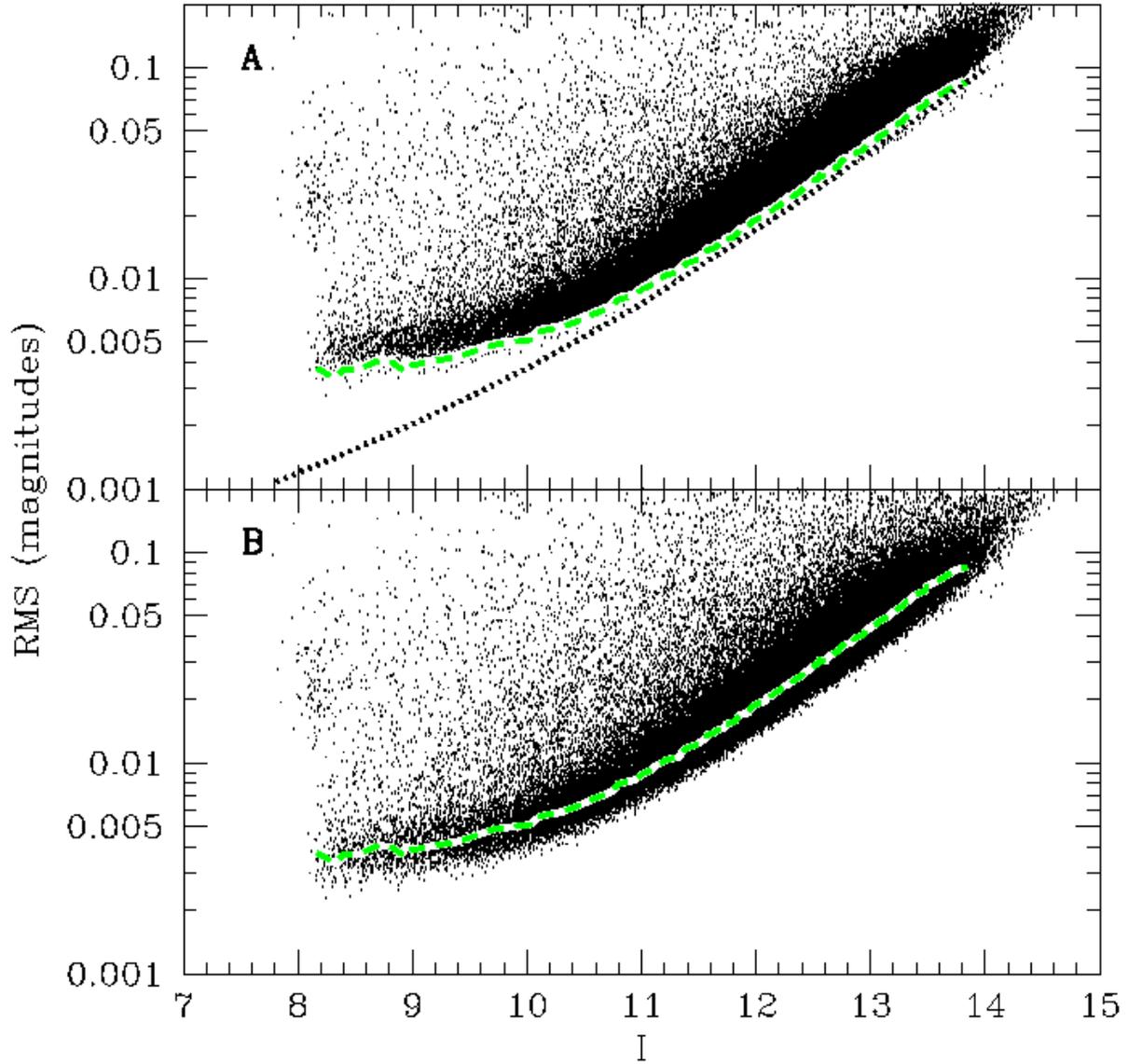}
\caption{(A) RMS vs $I$-band for the 97,540 lightcurves with RMS $<
0.2$ mag. The dashed line marks the lower
envelope of the plotted distribution. The dotted line shows the
theoretical limits (photon noise). (B) RMS vs $I$ for the light curves
binned by 20 minutes. The dashed line is the same as in (A). The fact
that binning reduces the RMS shows that the light curves are not
dominated by systematic errors that are correlated on time scales
longer than 20 minutes.}
\label{sig}
\end{figure*}

Figure \ref{sig} shows the light curve RMS vs.~reference magnitude for
our 98,000 objects. Because many of the light curves contain outliers
that result from non-random errors (e.g.~satellite crossings, bad
pixels, etc.), before calculating the RMS of Fig~\ref{sig}-A we first
sorted each light curve by magnitude and removed two outlier points
from each end. From this plot it is clear that we have obtained a
photometric precision down to 3.5 millimagnitudes at the bright
end. For the dim end the noise is dominated by the background and
rises to 0.01 magnitudes at I$\sim$11.3. This is very close to the
results that have been achieved on sparse stellar fields using
aperture photometry with the same instrument (see B04). From this we
conclude that we can achieve the same precision in a dense field with
image subtraction that we can in a sparse field with aperture
photometry. The fact that neither method is good down to the photon
limit implies that there is some as yet unexplained systematic error
that affects both methods. There are a total of 9,004 light curves
with RMS better than 10 millimagnitudes, and another 16,540 light
curves with RMS between 10 and 20 millimagnitudes.

Although we achieve very high precision for the brightest stars, it is
apparent that we are not photon-limited at the bright end. To determine
whether or not our systematic errors are correlated along the light
curves, we binned the light curves in time using a bin size of 20
minutes. This reduced the maximum number of points per light curve from
784 to 286, a factor of 2.7. If the errors were correlated on
time-scales longer than 20 minutes, this procedure would not affect the
RMS of the light curves. But for uncorrelated errors we would expect
the RMS to be reduced by 40\%. In Fig~\ref{sig}-B we show the
resulting sigma vs.~reference $I$ diagram, together with a line tracing
the bottom envelope from Fig~\ref{sig}-A. At the bright end the RMS
approaches 2.5 mmag, or a reduction of 29\%, which is consistent with the
errors being uncorrelated.

\subsection{Astrometry}

Because the ISIS package provides an image registration routine via
``interp.csh'', we do not follow the same steps as B04 to match images
to one another. Instead, once we have the X,Y positions of all point
sources in our field from DAOPHOT, we use the Delaunay-triangulation
algorithm as described in B04 to match to the Guide Star Catalog 2.2
(Bucciarelli et al.\ 2001).  Using the resulting RA/DEC grid for our
field, we match our entire list to {\it 2MASS}\/ using a matching
radius of $10\arcsec$. We then adopt RA/DEC from {\it 2MASS}\/ for the
matched sources.

Because {\it 2MASS}\/ contains many objects that are much fainter than
our upper magnitude limit, and is at higher resolution than our
observations, we only match to {\it 2MASS}\/ sources with $J<13$. Even
at this cut there are over 170,000 {\it 2MASS}\/ sources in the field
compared to our 98,000. Taking a cut at fainter magnitude will tend to
increase the number of spurious matches between our objects and fainter
{\it 2MASS}\/ sources. Although the magnitude that we measure for each
``object'' will be the summed $I$-band of all objects within roughly
$30\arcsec$, for the purposes of follow-up we will adopt the
convention that our ``object'' lies at the location of the nearest
{\it 2MASS}\/ source within $10\arcsec$ that has $J<13$.

Using a $10\arcsec$ matching radius we obtain 83,900 matches, with
6,174 objects having more than one {\it 2MASS}\/ source within
$10\arcsec$ (multiple matches). For all multiple matches, we choose
the closest match as the ``real'' one. We also matched with smaller
radii: for $1\arcsec$ there were 16,557 matches, and
for $5\arcsec$ there were 64,844 matches. To
determine whether or not the number of multiple matches is consistent
with random matching, we also shifted our entire starlist by
$15\arcsec$ and by $30\arcsec$ and matched it to {\it 2MASS}. For the
$15\arcsec$ shift we obtained 6776 matches, and
for the $30\arcsec$ shift we obtained 4103 matches. If the 170,000 {\it
2MASS}\/ objects were randomly distributed across the frame, one would
expect to find 1 object for every 24 pixels. Assuming that our 98,000
objects are also randomly distributed we would expect 6,400 random
matches between HAT and {\it 2MASS}. This is consistent with the number
of multiple matches that we see with the $10\arcsec$ matching radius,
and with the number of matches that we see with the $15\arcsec$ shift.
The fact that the number of matches drops as we shift to $30\arcsec$
may suggest that the objects are not randomly distributed but have some
degree of clustering.

There are a number of objects with $J>13$ that will not have matches
in our catalog. Using the distribution of colors $I-J$ obtained from
the {\it 2MASS}\/ matches with $10<I<11$ we estimate that roughly 5000
of the observed sources with $I>11$ should have $J>13$. The fact that
13,920 of the 14,100 unmatched objects have $I>11$ shows that most of
the unmatched objects ($\sim$9000) cannot be accounted for from the
{\it 2MASS}\/ cutoff. These sources may be spurious detections by
DAOPHOT, or they could be sources with bad astrometry. If we assume
that they are spurious then we can estimate that for $I\sim12.5$
approximately 10\% of the sources are spurious, for $I\sim13$
approximately 20\% are spurious, and for $I\sim14$ more than 30\% are
spurious.

With a pixel radius of $\sim15\arcsec$ we would expect to achieve
better than $1\arcsec$ precision in our astrometry. The fact that we
require a $10\arcsec$ matching radius is somewhat surprising. Indeed
in performing the magnitude zero-point calibration 90\% of the matches
to Hipparcos were better than $1\arcsec$. The Hipparcos stars were
bright ($I<10$), and we find that for our match to {\it 2MASS}\/ we
match to better than $1\arcsec$ for most of the bright stars. It seems
likely that there is a systematic error that affects the astrometry at
the faint end.

\section{Selection of Variables}

\subsection{Rescaling ISIS Errors}

\begin{figure}[p]
\epsscale{1}
\plotone{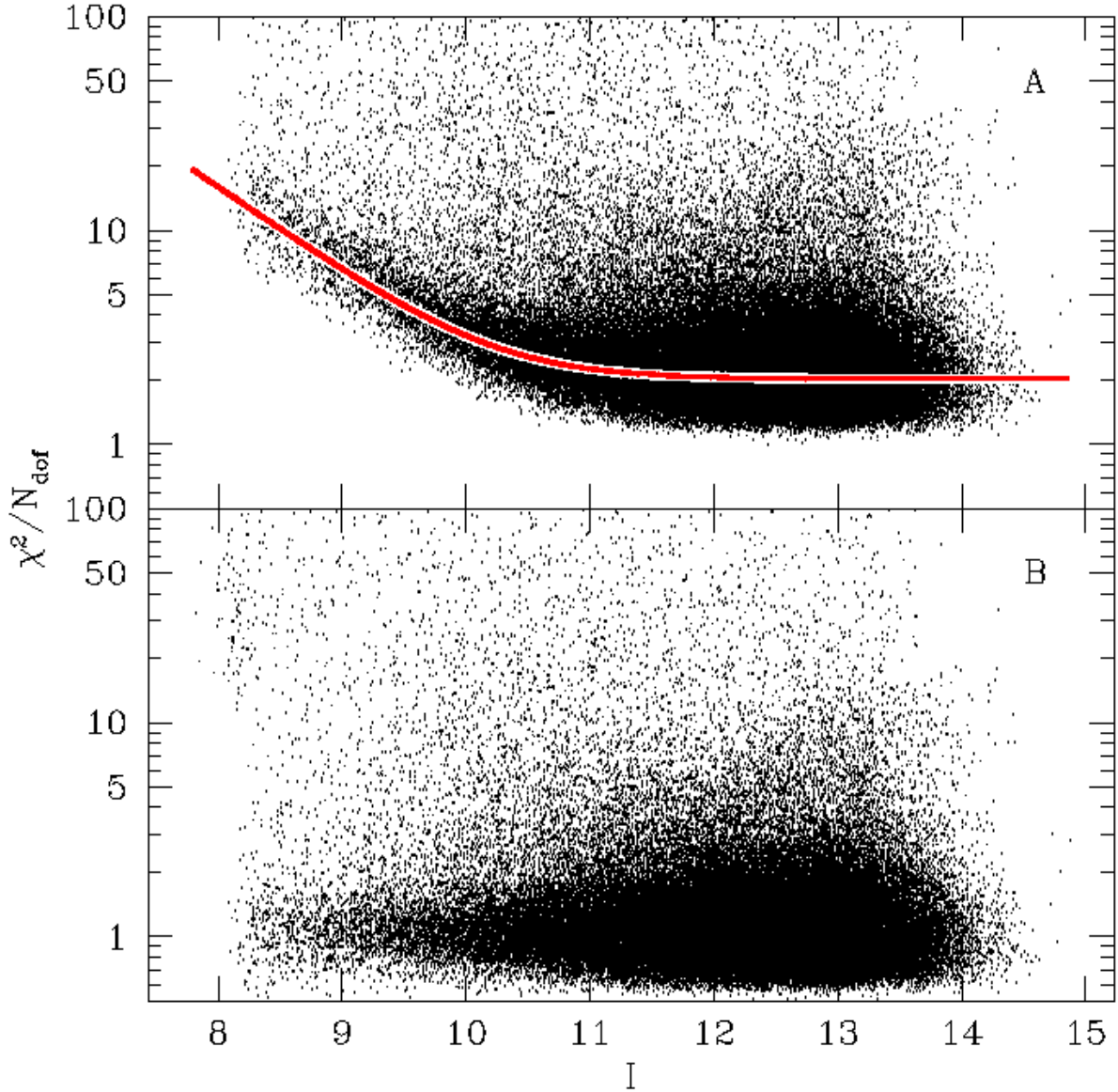}
\caption{(A) Reduced $\chi^{2}$ of 98,000 light curves vs. I magnitude
of the objects on the reference frame. The line shows the function
used to rescale the formal photometric errors from ISIS. (B) Reduced
$\chi^{2}$ vs. I following correction of formal flux errors from ISIS
(\S 4.1).}
\label{chisqrvI}
\end{figure}

Before proceeding with the selection of variables, it is useful to
rescale the formal flux errors from ISIS to match the empirically
observed errors. This is necessary since the formal errors are assumed
to represent the real errors when used by variability tests such as
Stetson's ``J'' (Stetson 1996). We follow a procedure similar to that
used by Kaluzny et al.\ (1998).

To do this we calculate the reduced chi-square 

\begin{equation}
(\chi^{2}/N_{dof})=\frac{1}{N-1}\sum_{i=1}^{N}\left
(\frac{I_{i}-\bar{I}}{\sigma_{i}}\right )^{2}
\end{equation}
for every light curve and plot it as a function of I
(Fig.~\ref{chisqrvI}). Here $I_{i}$ is the measured magnitude at time
$i$, $\bar{I}$ is the mean magnitude for the light curve, and
$\sigma_{i}$ is the formal error assigned to the magnitude measurement.
The observed $\chi^{2}/N_{dof}$ rises well above the expected value of
1 at the bright end. This is because systematic errors dominate the
light curves of the bright stars as described in \S 3.3. To account for
these systematics in our errors we fit a curve to the ``ridge'' of the
observed $\chi^{2}/N_{dof}$ vs.~$I$ distribution. We then multiply the
formal errors by the square-root of this function. The resulting
``corrected'' $\chi^{2}/N_{dof}$ vs.~$I$ is shown in
Fig.~\ref{chisqrvI}-B.

\subsection{Stetson's Variability Index}

As a preliminary selection of variable stars, we apply the Stetson
``J'' variability test (Stetson 1996). To apply this statistic one
forms $n$ pairs of observations each with a weight $w_{k}$ and then
calculates
\begin{equation}
J = \frac{\sum_{k=1}^{n}w_{k}{\rm
sgn}(P_{k})\sqrt{\left|P_{k}\right|}}{\sum_{k=1}^{n}w_{k}}
\end{equation}
where
$$P_{k}=\left\{
\begin {array} {l}
	\delta_{i(k)}\delta_{j(k)},{\rm~if~}i(k)\neq j(k),\\
	\delta_{i(k)}^{2}-1,{\rm~if~}i(k)=j(k)
\end{array}\right.
$$
is the product of the normalized magnitude residuals,
$$
\delta=\sqrt{\frac{n}{n-1}}\frac{I-\bar{I}}{\sigma_{i}}
$$
of the two paired observations. The pairing and weighting scheme that we use
is analogous to the one employed by Kaluzny et al.~(1998). We use a
time-scale of 30 minutes for pairing, assigning a weight of 1.0 to
pairs formed by distinct points, and a weight of 0.1 to ``pairs''
formed from a single point.

\begin{figure}[p]
\epsscale{1.0}
\plotone{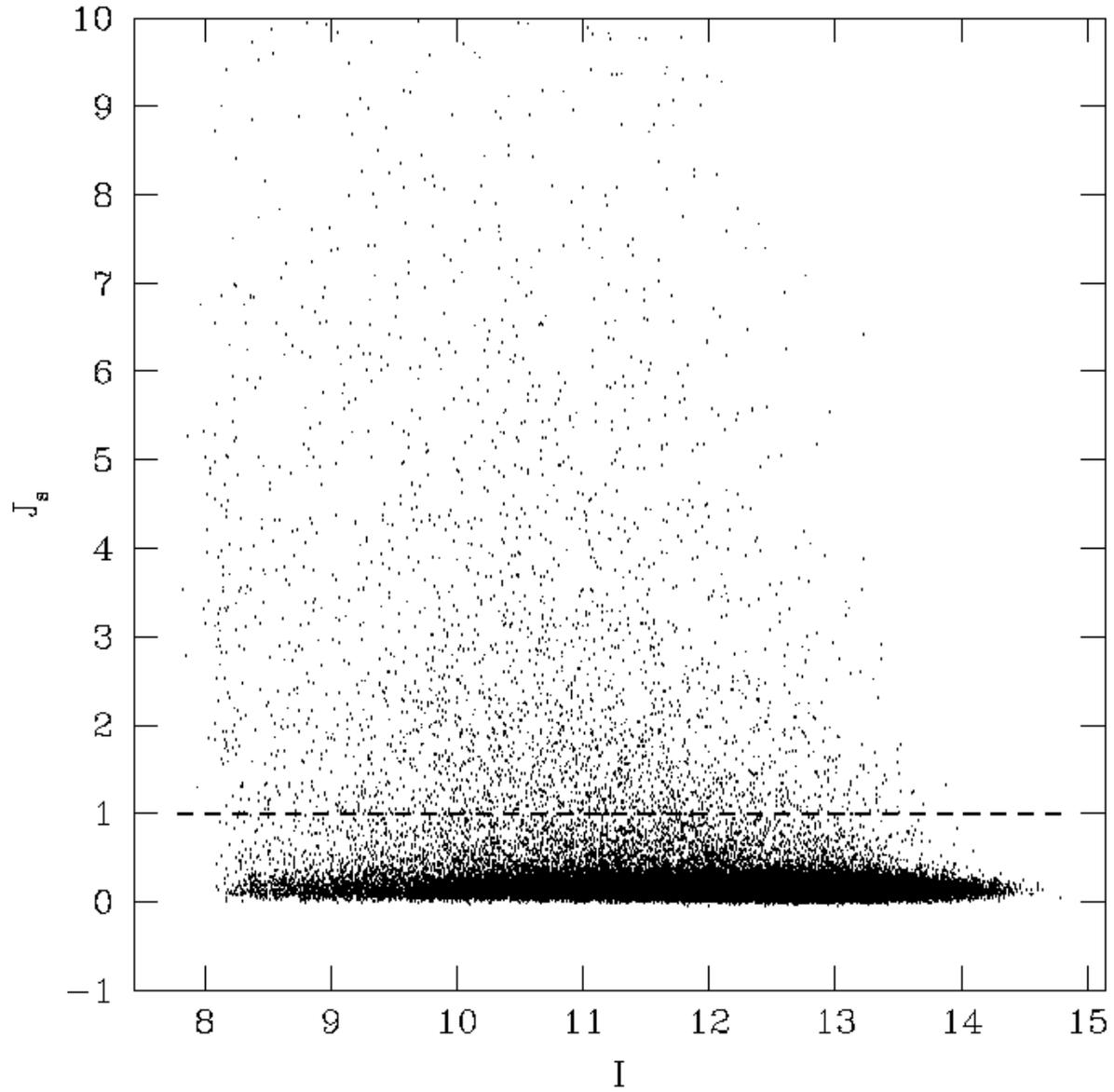}
\caption{Plot of Stetson's ``J'' variability index ($J_{s}$) vs I
magnitude. The 2830 objects lying above the dashed line ($J_{s}=1.0$)
were selected as candidate variable stars (\S 4.2).}
\label{Jstetson}
\end{figure}

To select the variable stars we apply a cut of $J_{s}>1.0$
(Fig.~\ref{Jstetson}). This selects 2830 light curves, all of which
show some form of correlated variability.

\subsection{Selection of LPVs}

The majority of light curves selected with the above procedure do not
show any periodicity within the 30 day window of observations.
Typically these light curves increase or decrease monotonically over
the run, although there are some that achieve a minimum or maximum
magnitude. For the purposes of this paper we define a ``Long Period
Variable'' (LPV) to be any variable for which the fit to a parabola is
substantially better than the fit to the mean. These are most likely
Mira variables or semi-regular/irregular variables.

\begin{figure}[p]
\epsscale{1}
\plotone{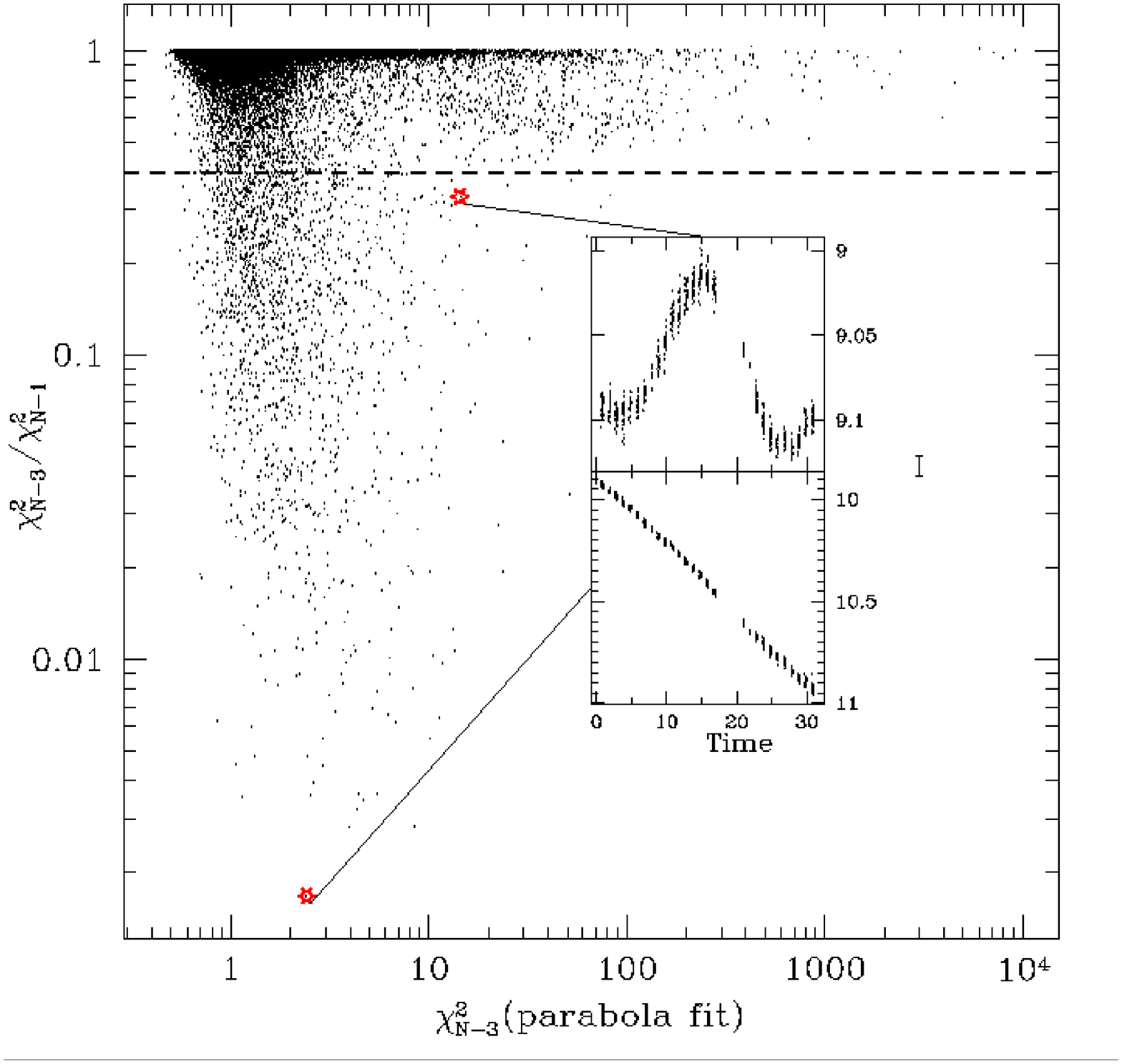}
\caption{Ratio of reduced chi-square for fit of parabola to reduced
chi-square about the mean ($\chi_{N-3}^{2}/\chi_{N-1}^{2}$) vs reduced
chi-square for fit of parabola $\chi_{N-3}^{2}$. This plot is for all
98,000 light curves. The 1535 objects below the dashed line
($\chi_{N-3}^{2}/\chi_{N-1}^{2}=0.4$) that also had $J_{s}>1.0$ are light
curves that we classified as LPVs (\S 4.3). (Inset) Light curves with
$\chi_{N-3}^{2}/\chi_{N-1}^{2}\ll 0.4$ are better fit with a parabola than
those with $\chi_{N-3}^{2}/\chi_{N-1}^{2} \lesssim 0.4$.}
\label{lpvchisqr}
\end{figure}

To separate the LPVs from the other variable stars we apply the
following simple cut. We first fit a parabola to all the light curves
flagged as variable by the $J_{s}$ cut. We then calculate the reduced
chi-squared, $\chi^{2}/N_{dof}$ (we use the short-hand
$\chi_{N_{dof}}^{2}$, e.g. $\chi_{N-3}^{2}$ in the case of fitting to
a parabola), for each fit. True LPVs will show dramatic improvement
when fit with a parabola as opposed to fitting with the mean
($\chi_{N-1}^{2}$). To select the LPVs we perform an F-test,
classifying any light curve with $\chi_{N-3}^{2}/\chi_{N-1}^{2} < 0.4$
as an LPV (Fig.~\ref{lpvchisqr}). This procedure selects 1535
candidate LPVs, leaving 1295 candidate non-LPV variables.

We chose this cut empirically at a point where light curves that may be
best fit with a 3rd order polynomial begin to be mixed in with the
parabolic light curves. An example of how a light curve with
$\chi_{N-3}^{2}/\chi_{N-1}^{2}\ll 0.4$ compares to a light curve with
$\chi_{N-3}^{2}/\chi_{N-1}^{2} \lesssim 0.4$ is shown in the inset of
Fig.~\ref{lpvchisqr}.

\subsection{Removal of Spurious Variables}

\begin{figure}[p]
\epsscale{1}
\plotone{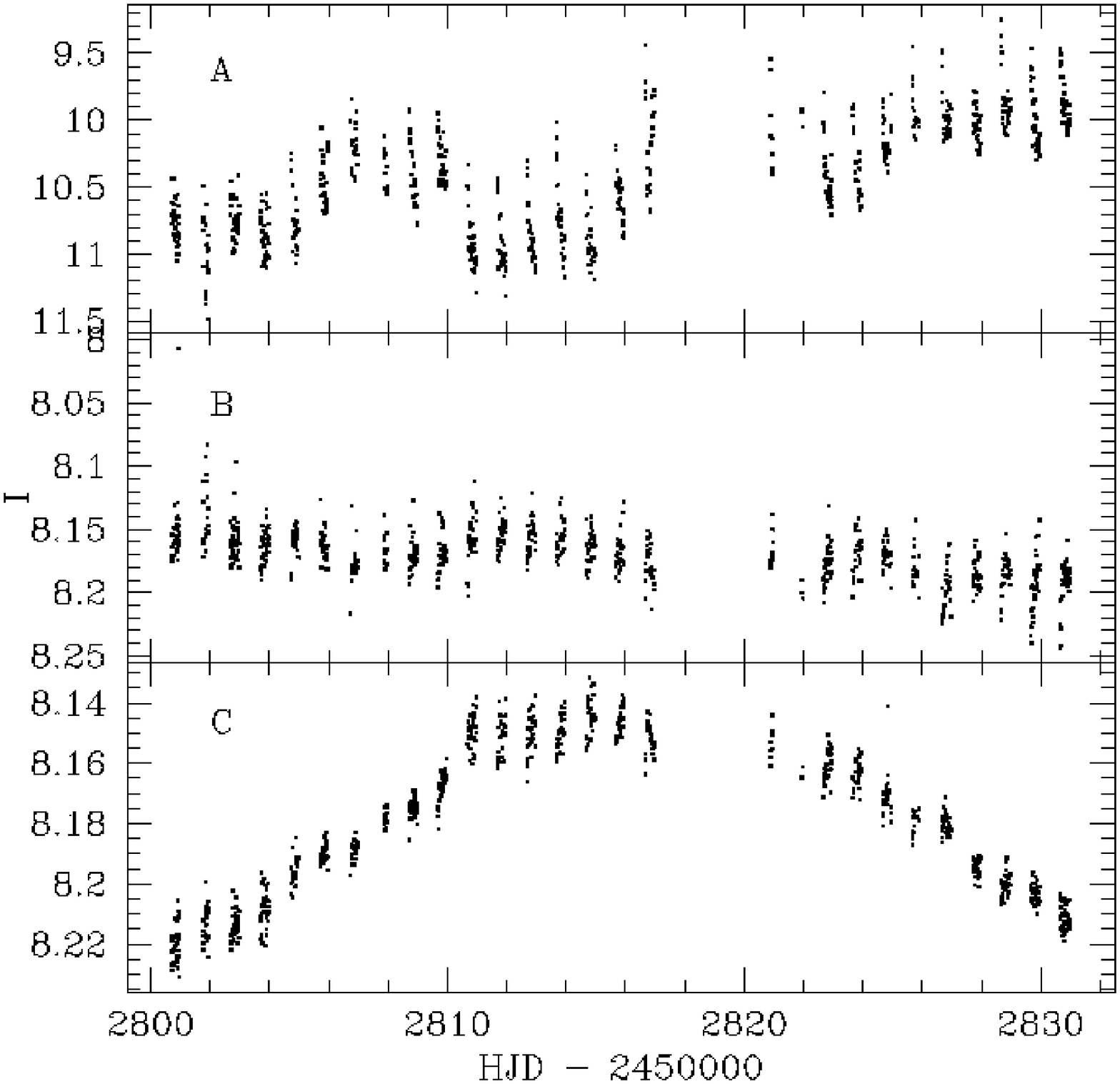}
\caption{Light curves used as templates to remove spurious variables 
(\S4.4).}
\label{trends}
\end{figure}

In \S3.3 we claimed that our systematic errors are uncorrelated on
time-scales longer than 20 minutes. Although this is true for the
majority of stars, we do see a number of light curves that show nearly
identical variations. These trends can be classified into three basic
types: template1-like (Fig~\ref{trends}-A), template2-like
(Fig~\ref{trends}-B), and template3-like (Fig~\ref{trends}-C). Since
image subtraction assumes conservation of flux, there are a number of
light curves that show the same trends, but reflected about the
horizontal axis. 

Regardless of the nature of these systematic variations, we believe
the most straightforward way to reject light curves that show these
trends is to reject those that have a substantially smaller
chi-squared when fit, point-by-point, to the template light curves,
than when fit to the mean. When fitting to the templates we allow one
free parameter: an overall scaling of the variations above and below
the mean. We apply different cuts for the LPV and non-LPV light
curves, since a very strict cut tends to reject more LPVs than
non-LPVs, particularly for fitting to template1. For template3 we also
imposed a magnitude cut to avoid removing variables that had a
numerically good fit, but had a maximum at a different time and were
thus probably not spurious. Looking back through the permitted LPVs we
rejected 4 candidates by eye that showed a strong resemblance to
template3 or template2. The template3 cut was applied after all other
cleaning steps. We also observed a number of light curves that
appeared to be correlated with the airmass of the observations. These
tended to be removed more efficiently by fitting to template1 or
template2 than by defining a separate template, as a number of
``true'' variables showed small 1 day oscillations on top of an
over-all ``real'' variation, and would be rejected by a separate
template.

As a check on our template3 rejections we examined the light curves
released by the Northern Sky Variability Survey (NSVS, Wozniak et al.\
2004) for the 13 objects rejected by template3. The survey made use of
the Robotic Optical Transient Search Experiment (ROTSE-I) to provide
to the public a temporal record of the northern sky over the optical
magnitude range from 8 to 15.5. Notably the optics and CCD used for
this telescope are now in use on the HAT-5 instrument. The majority of
our template3-like light curves have a full-amplitude of $\sim 0.05$
mag, however as they are classified as LPVs, one might expect that if
they are real then a number of them should have full-amplitudes
greater then $\sim 0.1$ mag when observed over a longer baseline. The
typical RMS for the NSVS light curves of these objects is $~0.05$ mag,
so any variations with full-amplitude less than $\sim 0.1$ mag would
be unrecognizable in these light curves. We find that 3 of the
template3-like light curves appear to show variations in the NSVS.
These 3, including HAT199-648, HAT199-3997 and HAT199-4205 will remain
in our catalog. It should be noted that all three of these light
curves show an inverted template3-like shape, and that 3 of them have
full-amplitudes greater than 0.1 magnitudes in our observations, and
are therefore already suspect as template3-like.

For the variables classified as LPVs we reject 9 light curves with
$\chi^{2}_{temp1}/\chi_{N-1}^{2}<0.35$, 10 light curves with both
$\chi^{2}_{temp3}/\chi_{N-1}^{2}<0.23$ and $I_{ref}<10.0$, and an
additional 4 by eye. For non-LPVs we reject 359 light curves with
$\chi^{2}_{temp1}/\chi_{N-1}^{2}<0.6$ and another 99 light curves with
$\chi^{2}_{temp2}/\chi_{N-1}^{2}<0.85$.

As a further cleaning step we also rejected 46 of the non-LPV light
curves that had fewer than 693 points. We did not reject 2 light curves
that had fewer than 693 points but appeared to show real variability.

\subsection{Selection of Periodic Variables}

To search the remaining 791 candidate non-LPV variable stars for
periodicity, we use a variation of the period finding algorithm by
Schwarzenberg-Czerny (1996). This algorithm is implemented in a code
due to J.~Devor (private communication). The codes provides the two
``optimal'' periods, along with a measure of confidence for these
periods ($\sigma_{AoV}$).

\begin{figure}[p]
\epsscale{1}
\plotone{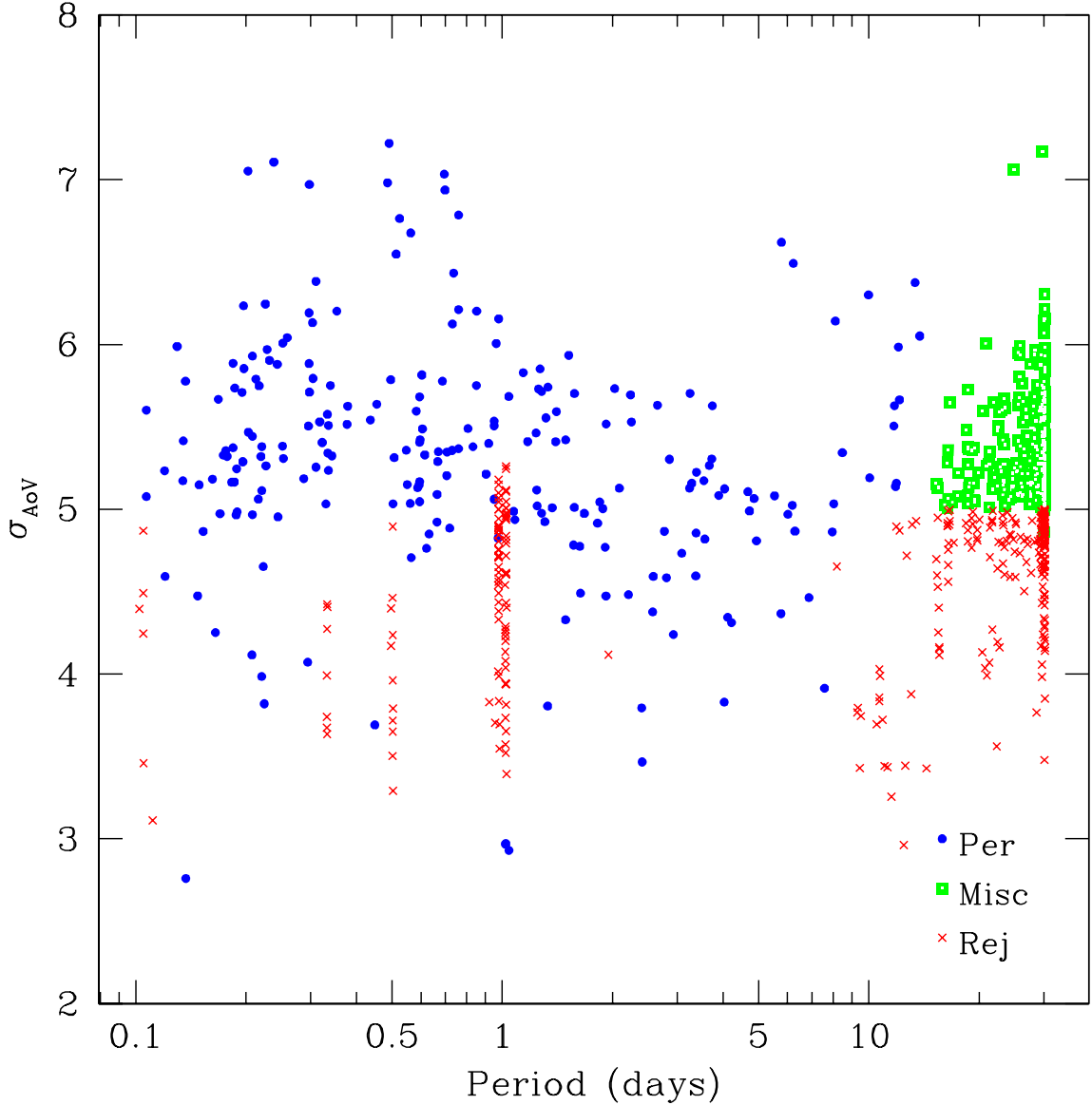}
\caption{Period vs $\sigma_{AoV}$ for the remaining 791 non-LPV
variables. $\sigma_{AoV}$ is the measure of confidence assigned to the
period fit (see \S 4.5). The dots show the 231 objects classified as
``periodic'', the boxes are the 266 objects classified as
``miscellaneous'' and the x's are the 294 rejected light curves.
Periodic light curves have periods less than 14 days and do not (except
in a few cases) have periods near a harmonic of one day. Miscellaneous
light curves have $P>14$ days and $\sigma_{AoV}>5$.}
\label{Psig}
\end{figure}

Using the best period as a starting point, we proceeded to classify the
remaining 791 non-LPV light curves by hand as either periodic (with
period less than 14 days), miscellaneous (a light curve that is not an
LPV, and does not have a period $<$ 14 days) or a light curve to reject
(typically light curves that appeared to be dominated by periods that
were harmonics of one day, or light curves that resembled the trends of
\S4.4). We chose a 14 day cut-off for the period to ensure that any
light curve we classified as periodic completed more than 2 full
periods within the window of observations. Since rejecting light
curves by eye is a highly subjective procedure we decided to find
objective cuts that would generally yield the same subjective
classifications. We rejected light curves whose best period fell near
a harmonic of one day and had a low value of $\sigma_{AoV}$, or light
curves with a best period greater than 8 days and $\sigma_{AoV} <
5$. For the remaining light curves we called the object ``periodic''
if the best period was less than 14 days, and ``miscellaneous'' if
not. These particular cuts were chosen empirically as providing the
cleanest removal of ``suspicious'' light curves. The results of these
cuts are shown in Figure \ref{Psig}. We looked through the rejected
light curves and rescued 8 cases that we believed were clearly
variables. We also looked through the periodic and miscellaneous light
curves, rejecting 5 light curves that showed a significant resemblance
to one of the templates (\S4.4). There were 294 light curves rejected
in this step. We erred on the side of caution for the rejection cuts,
a look through the rejected light curves reveals many other probable
``true'' variables. For this reason we do not claim completeness for
our periodic or miscellaneous class variables.

\subsection{Blending}

Following the above classification/cleaning procedures, and prior to
the removal of the template3-like light curves, we were left with 3
sets of variable stars: 1526 LPVs, 266 miscellaneous variables, and
231 periodic variables. However, as mentioned in \S 3.2, one
short-coming of the ISIS photometry program ``phot.csh'' is that it
does not account for blending in the subtracted images. Although true
variables are generally well separated in the subtracted images, any
given variable may have a number of non-variable stars within a {\tt
rad\_aper}. As a result, ``phot.csh'' will sample a portion of the
variable flux from the nearby variable star when measuring the
magnitude of the non-variable star, and our procedure may flag some of
these stars as variables. Existing routines, such as DAOPHOT, that
perform ``de-blending'' profile photometry do not allow for negative
fluxes, and as such will not work on subtracted images. Moreover,
running these routines on the absolute value of the subtracted images
will have difficulty dealing with the case of two nearby variables,
with one a positive variation and the other a negative
variation. Since ``phot.csh'' weights each pixel in the aperture by
the PSF, ``phot.csh'' will measure less flux for the non-variable
stars than for the true variables, as the non-variables will be
off-center from the variable flux. This suggests a method to separate
the true variables from the blended light curves.

The first step is to identify blending groups. From each blending group
we identify the true variable as the light curve that has the highest
standard deviation (in flux). To select the blending groups we first
find all pairs of variable stars that are separated by fewer than 6
pixels. There are a total of 461 pairs (well above the expected number
due to random matching) involving 695 distinct light curves. We then
form groups so that pairs like 1-2 and 2-3 will be grouped into 1-2-3
etc. The largest groups contained 5 light curves (there were 3 of these
groups). This formed 305 groups so that 390 objects were rejected as
blended light curves. Two of the remaining periodic variables were then
removed by hand as they showed no obvious periodicity, and a strong
likeness to template1.

Following the correction for blending (and removal of template3-like
LPVs as per \S 4.4) we arrived at our final list of variables
consisting of 1169 LPVs, 241 miscellaneous variables and 207 periodic
variables for a total of 1617 distinct variables. The periodic
variables were then classified, by hand, into two general classes:
eclipsing variables (157 objects) and pulsating variables (50 objects).
These further classifications are subjective and represent the authors'
suspicions as to whether or not the light curve appears to show some
form of eclipses.

Because we only choose one true variable from each blending group we
will reject true variables that lie within 6 pixels of other, larger
amplitude (in flux), true variables. When two true variables are nearby
one another, the light curves of both objects will likely show
variability blending. Our procedure will reject the true variable whose
flux light curve has a smaller standard deviation. However, the light
curve of the accepted variable will likely show some variations due to
the rejected variable. For 1617 variables randomly distributed across
the image, we expect $\sim$ 70 such pairings, or roughly 4\% of all cases.
Since we are more inclined toward correctly identifying variables than
forming a complete list (that will include many false-positives) we
choose to live with the rejections. It should be noted, however, that
as many as 4\% of our variable star light curves may show contamination
from another nearby variable star which is not included in the list.
This number may be even higher if there is clustering as suggested in
\S3.4.

\begin{figure}[p]
\epsscale{1}
\plotone{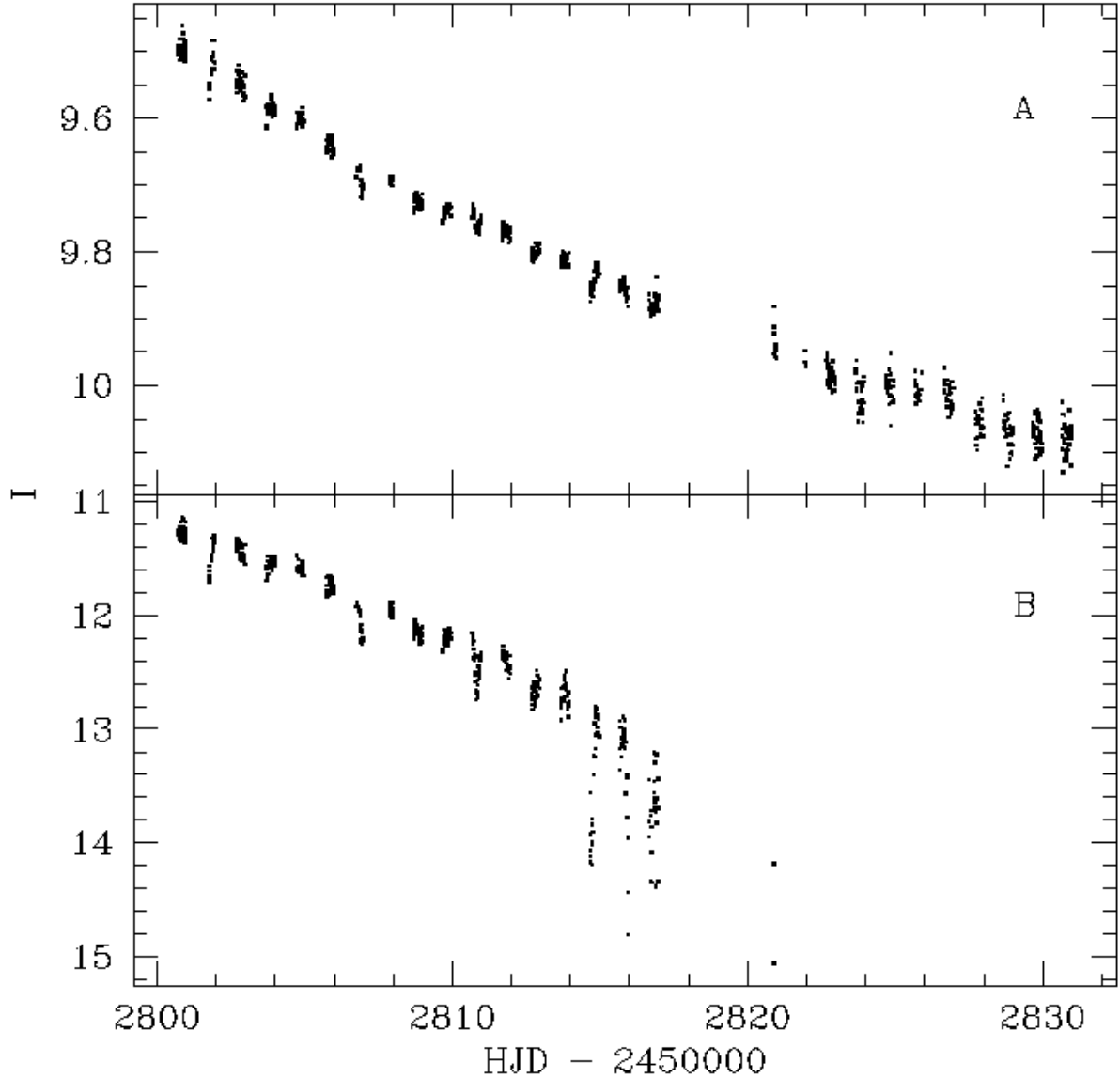}
\caption{(A) LPV matched to V484 Cyg, an EA/SD binary, as well as to
V1360 Cyg, a Mira variable. The eclipses are less pronounced than in
(B), meaning that this light curve likely corresponds to the Mira V1360
Cyg. (B) Light curve of an object within 2 pixels of (A). This light
curve shows deeper eclipses than (A) and is likely V484 Cyg. The larger
flux variations due to (A) are blended into this light curve. V484 Cyg
is not included in the catalog of variables.}
\label{lcblend}
\end{figure}

As an example of this variability blending consider
Figure~\ref{lcblend}. Figure~\ref{lcblend}-A shows the light curve of
an object that we matched with V484 Cyg, an EA/SD binary with eclipses
of 1 mag in V (between 13.5 and 14.5) and a period of 1.29 days, as
well as to V1360 Cyg, a known Mira variable. Our observations reveal a
monotonic decrease in flux from $I=9.48$ to $I=10.11$ over 30 days,
with slight eclipses. A search of nearby objects revealed a star within
2 pixels with a light curve that showed deeper eclipses on top of an
overall declining envelope (Fig~\ref{lcblend}-B). Indeed both these
objects are matched to separate {\it 2MASS}\/ objects. This, we
believe, is an especially pronounced case of variability blending where
two real variables lie within each others' aperture and hence the
variability is blended into both objects. In this case
Fig~\ref{lcblend}\/-B likely corresponds to V484 Cyg, and
Fig~\ref{lcblend}-A to V1360 Cyg. Our selection method retained V1360
Cyg while rejecting V484 Cyg as a blended light curve. Because the
light curve of V484 Cyg is strongly corrupted by the presence of the
nearby LPV, we will not attempt to ``rescue'' it into the catalog, and
will only include V1360 Cyg.

It is important to stress that in this procedure we do not, in any
way, correct for the blending that results from resolution limits
whereby a number of distinct sources are blended into an individual
object. Indeed many of the individual light curves may consist of the
summed light from several sources lying along the same line of
sight. Instead, the blending that we correct for is the blending of
variability into multiple resolved objects that results from ``phot.csh''
applying simple weighted-aperture photometry on the subtracted images.

\section{Catalog of Variables}

The full catalog of variable stars, including {\it 2MASS}\/
coordinates and IDs where available, will be available with the
electronic version of the refereed paper. The catalog and light curves
are also available on the
world-wide-web\footnote{http://cfa-www.harvard.edu/~gbakos/HAT/LC/199/}. For
illustration we display the first 9 rows of the catalog in
Table~\ref{tab1} and Table~\ref{tab2}. In the table we provide the
HAT-ID for each object, which uses the form HAT199-?????, where the
number is between 00001 and 98000 and sorts the light curves by
reference magnitude. In this section we discuss the overall properties
of the various classes of variables, as well as some of the
interesting cases from each class.

\subsection{Matching to Known Variables}

To cross-check with known sources, we matched our list of variables to
the Combined General Catalog of Variable Stars ({\it GCVS}, Kholopov
et al.\ 1998). The {\it GCVS}\/ contains 334 objects in our field; we
obtained matches to 159 of these using a 30\arcsec~matching radius. As
mentioned above, one of our sources matched to two independent {\it
GCVS}\/ sources, so that only 158 of our sources were classified in
the {\it GCVS}. To match we used {\it 2MASS}\/ coordinates for our
objects where available. There are 82 matches that lie within
5\arcsec. We take a liberal matching radius 30\arcsec~to allow for
matches to variables that do not have {\it 2MASS}\/ coordinates, as
well as to allow for the possibility that some of our variables are
matched to the incorrect {\it 2MASS}\/ counterpart. We also note that
the positions in the {\it GCVS}\/ catalog may come from a wide variety
of epochs, further necessitating the liberal matching radius.

Of the 176 {\it GCVS}\/ variables that do not match with one of our
variables, 111 of these have $V>13.5$ and are thus likely to either be
too faint to detect as stars, or so faint that the variations are lost
in the background noise in our observations. There are 24 unmatched
variables with $V<10$ that appear to be correlated with saturated
stars in our I-band observations. Of the remaining 40, there are 9
eclipsing binaries, 7 of which are algol-like (EA), and could go
undetected if none of the eclipses are observed, or if only small
portions of a few eclipses are observed. There are 8 Miras for which
the change in magnitude over 30 days may have been too small to
detect. Of these 8, the 5 that have ephemerides available in the {\it
GCVS}\/ appear to have been too dim to detect during the observations,
or varied by an amount less than 10 $\sigma$ above the noise. There
are 19 semi-regular and slow irregular variables, all of which may
have only varied slightly over the course of our observation. Finally
we can expect $\sim5$ {\it GCVS}\/ variables to be excluded from the
catalog due to variability blending. We see that we can account for
all of the {\it GCVS}\/ variables in our field to which we did not
obtain matches.

We also matched our catalog to the New Catalogue of Suspected Variable
Stars, including the supplemental series ({\it NSV}, Kukarkin et al.\
1982). Using the same matching radius as above, we obtained matches to
20 sources. The {\it NSV} designation for these confirmed variables is
provided in our catalog.

\begin{figure}[p]
\epsscale{1}
\plotone{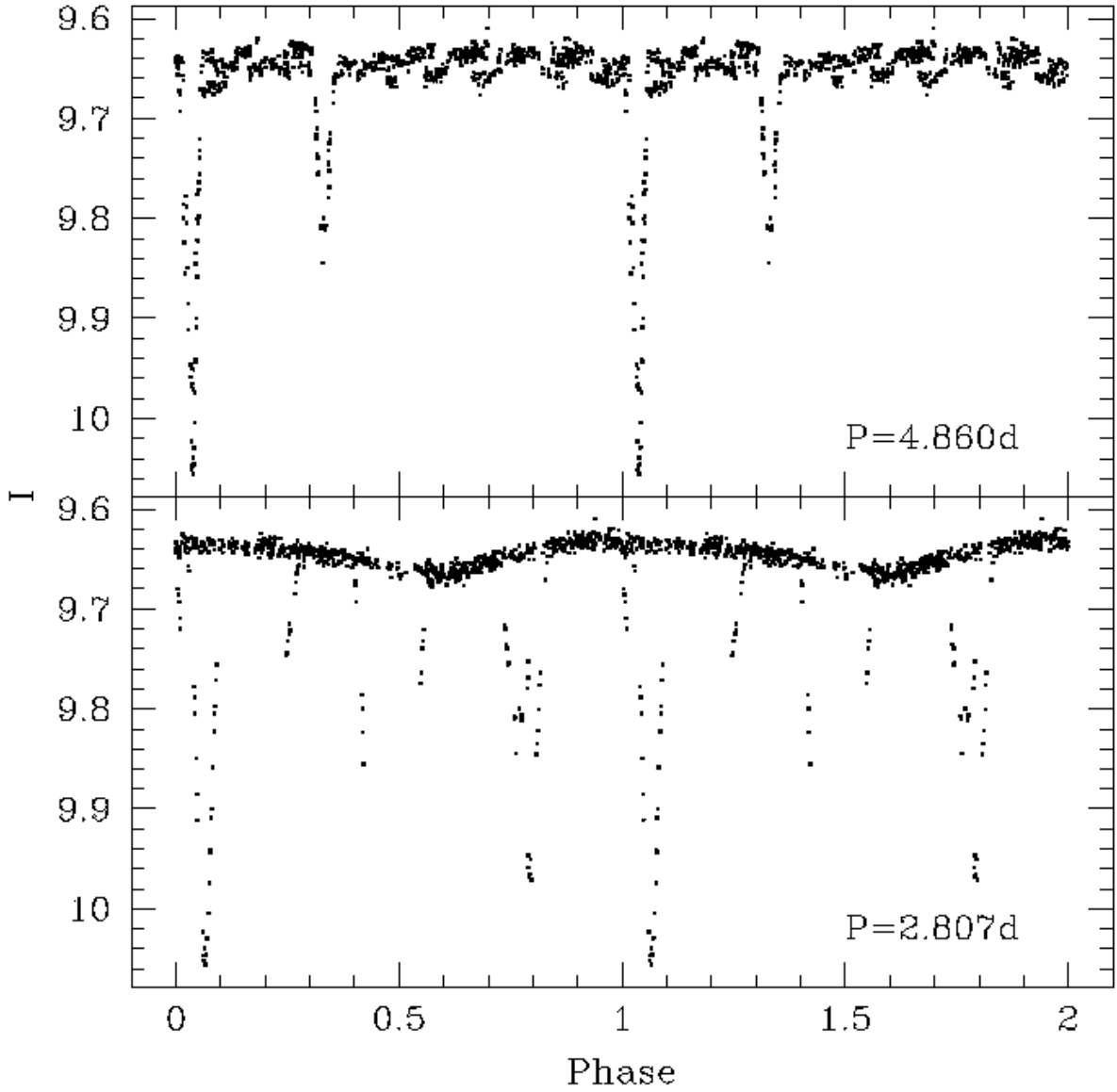}
\caption{Phase folded light curve of HD227269 shows eclipses with a 
period of 4.860 days and pulsations with a period of 2.807 days. The
system may be a blend between two distinct variables separated by
$\sim$9\arcsec~(\S5.1).}
\label{lc2743_mag.data}
\end{figure}

There have also been some more recent searches for variability in
fields overlapping this one. Notably Alonso et al.\ (2003), as part of
the STARE project, found that over 40 of the $\sim14000$ stars observed
in their Cyg0 field had pulsation periods between 5 and 40 days. Of
these they identify HD227269 as a highly eccentric eclipsing binary
showing possible pulsations. We do detect HD227269 as an eclipsing
binary, and also see the same pulsations (Fig.~\ref{lc2743_mag.data}).
Alonso et al.~mention, however, that a DSS image of the star reveals a
companion that is similar in brightness within $\sim$9\arcsec. At this
separation the stars would be blended in our observations, and it is
possible that the pulsations are not occurring in the binary system,
though this system does merit further investigation.

Another transit search that we overlap with is the Vulcan Photometer
project which has found over 50 eclipsing binaries out of 6000 stars
observed (Borucki et al.\ 2001). Of the brightest 6000 stars we
observed, 24 were eclipsing binaries. However, because our brightest
stars may contain many more blended objects than the 6000 relatively
isolated stars observed by Vulcan, these two populations may not be
directly comparable. The differences in the detection rates may also
be partially explained by differences in classifications. We should
also note that the Vulcan binaries include many low-amplitude systems
which may not have $J_{s} > 1$ and hence would not be flagged as
``large-amplitude'' variables by our method.

In \S4.4 we made use of the NSVS light curve database (Wozniak et al.\
2004) to manually check the rejected template3-like light curves to
determine if any of these showed variability in an independent
experiment. It may be useful in the future to also compare our entire
catalog of variables with the NSVS database, however that is beyond the
scope of this paper.

\subsection{Long Period Variables}

\begin{figure}[p]
\epsscale{1.0}
\plotone{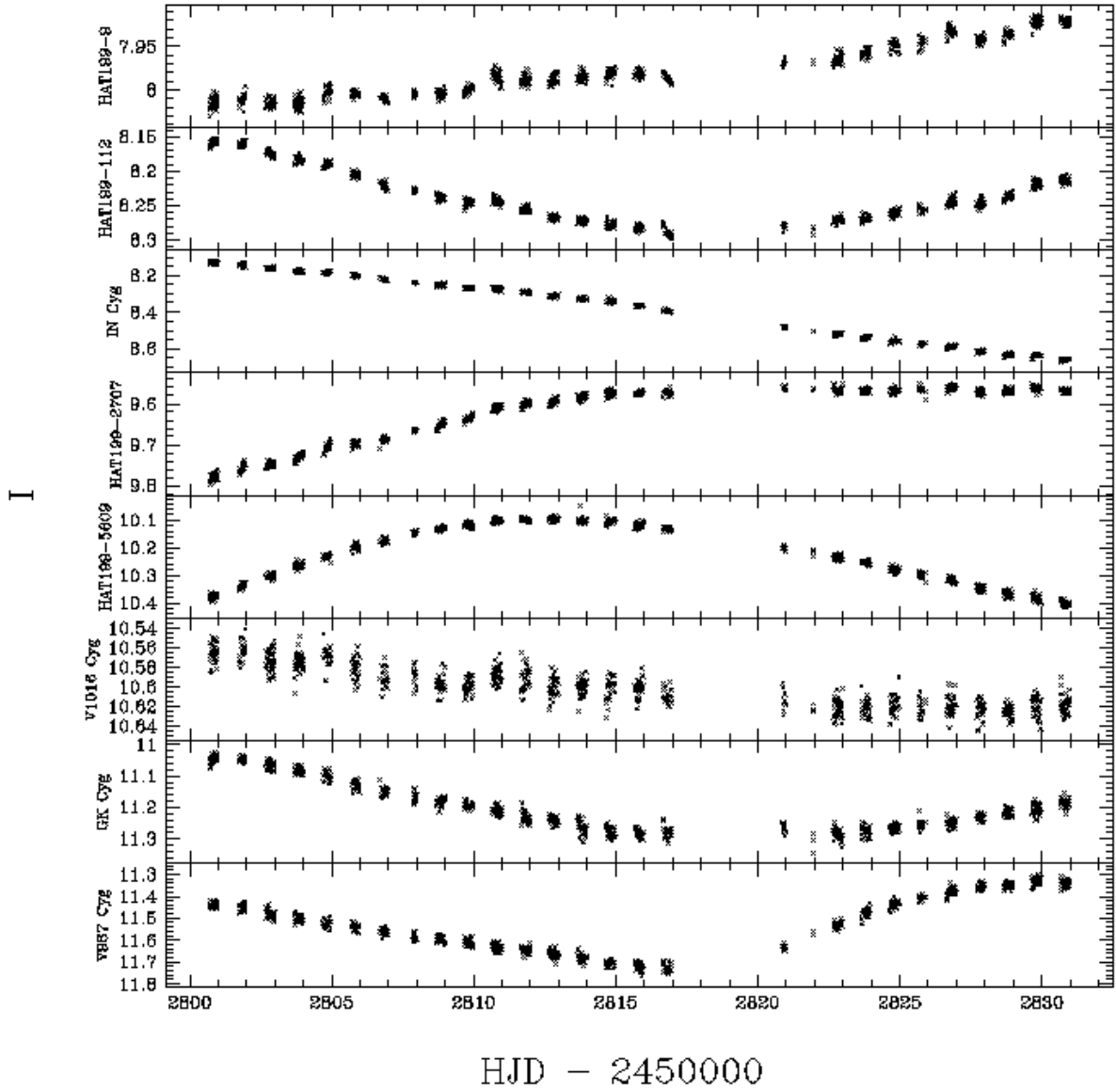}
\caption{I-band light curves of 8 LPVs in our catalog. A few of the 
variables that matched with the {\it GCVS}\/ include: V1016 Cyg, a
symbiotic nova, and GK Cyg and V967 Cyg which are RV Tau variables. }
\label{LPVfig}
\end{figure}

As mentioned in \S 4.3, we cataloged any object whose light curve was
well-fit by a 2nd order polynomial as an LPV. We identified 1169 LPVs,
of which 1026 are newly discovered variables. Of the known variables,
19 are newly confirmed {\it NSV} sources. Figure \ref{LPVfig} shows a
few of the interesting cases.

The majority of the LPVs are likely to be Mira variables which all have
periods longer than our 30 day window. Indeed 77 of the 97 matched LPVs
are known Miras. Recently there has been a great deal of interest in
pinning down the various P-L relations for Miras observed by the
microlensing surveys (e.g.~Wood et al.\ 1999, Wood\ 2000, and most
recently Groenewegen\ 2004). All of these have used observations of the
Large and Small Magellanic Clouds where the distances can be factored
out of the relations. Because we cannot assume a uniform distance for
our observations, this population of Miras will likely be less useful
towards this endeavor. Our population, however, is substantially
brighter and hence may be more useful for detailed investigations of
AGB stars.

\begin{figure}[p]
\epsscale{1}
\plotone{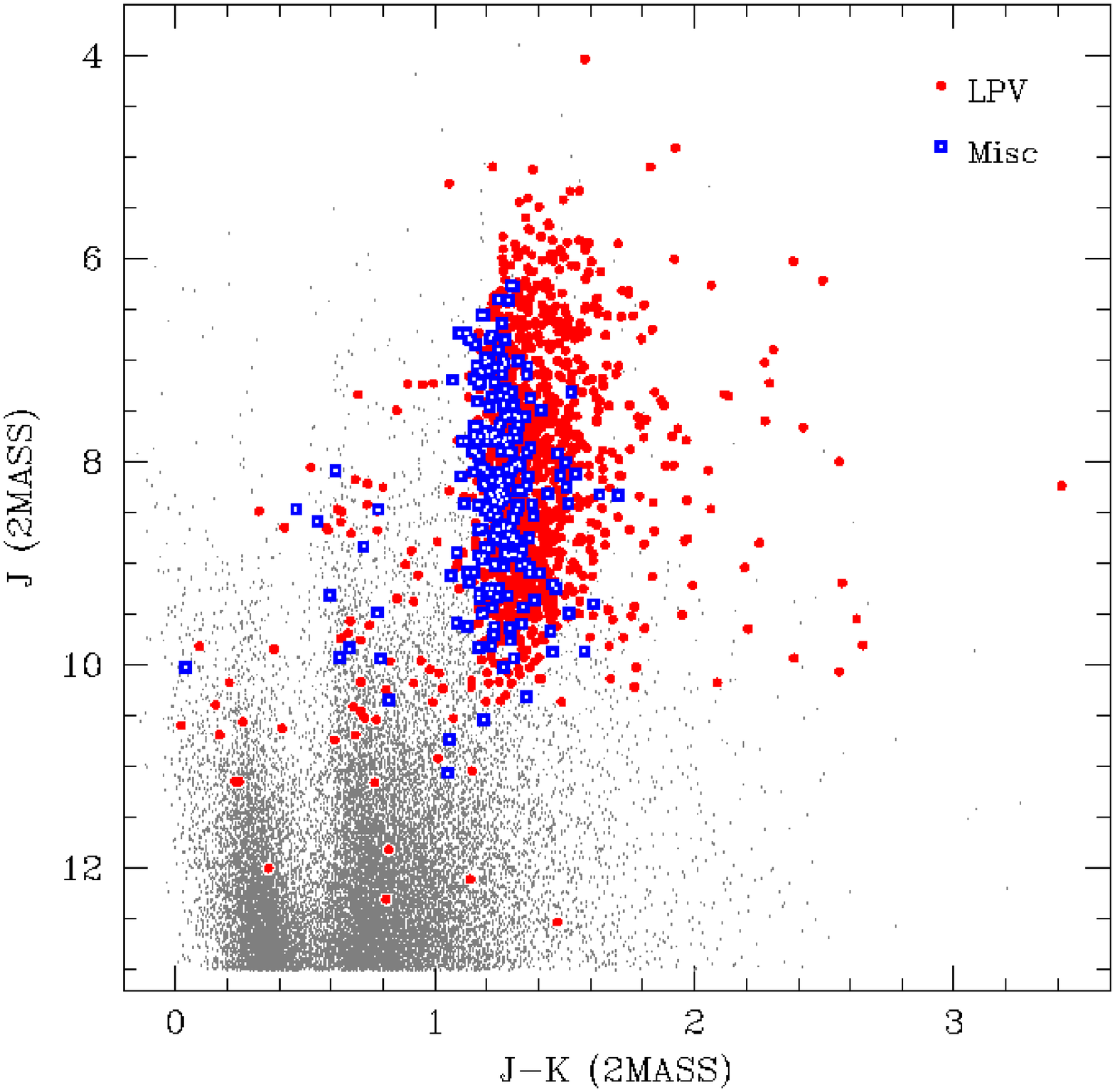}
\caption{J vs. J-K CMD showing location of LPVs and miscellaneous
variables relative to all {\it 2MASS}\/ objects in our field with
$J<13$. Magnitudes are taken from {\it 2MASS}. LPVs are shown as dots,
miscellaneous variables as boxes, and the general population of objects
is shown in grey. Note that both the LPVs and miscellaneous variables
are generally redder than the overall population with LPVs being
typically redder than miscellaneous variables. LPVs are mainly Mira
variables, while miscellaneous may include many type I and type II
Cepheids that have periods between 14 and 30 days. The bluest
miscellaneous variable, with $J=10.02$ and $J-K=0.039$ is V1920 Cyg, a
PV Tel type variable (\S 5.4).}
\label{cmd1}
\end{figure}

Figure~\ref{cmd1} shows the location of the LPVs on a J vs. J-K color
magnitude diagram (CMD). The infrared magnitudes J, and K are taken
from the match to {\it 2MASS}. As expected the LPVs are generally
redder than the majority of stars, and tend to lie along the giant
branch of the CMD.

Among the more exotic variables that we classify as LPVs are V1016 Cyg,
a well-studied symbiotic nova whose cool component is a $P=474$ day
Mira (e.g.~see Parimucha\ 2003), and a few RV Tau stars including GK
Cyg, and V967 Cyg.

\subsection{Periodic Variables}
We identified 207 large amplitude (full amplitude greater than 0.032
mag) variables that show periods less than 14 days; 180 of these are
newly discovered. As discussed in \S4.5 this cutoff at 14 days was to
ensure that any star classified as periodic had been observed for two
full periods. We further classified the periodic light curves into 157
eclipsing binary-like (EB) light curves, and 50 pulsating
variable-like light curves. Figure~\ref{EBfig} shows light curves for
48 of the EBs, and Figure~\ref{PULfig} shows 48 of the pulsating
variables (the other two pulsating variables are shown in
Figure~\ref{multiperiod}).

\begin{figure}[p]
\epsscale{1.0}
\plotone{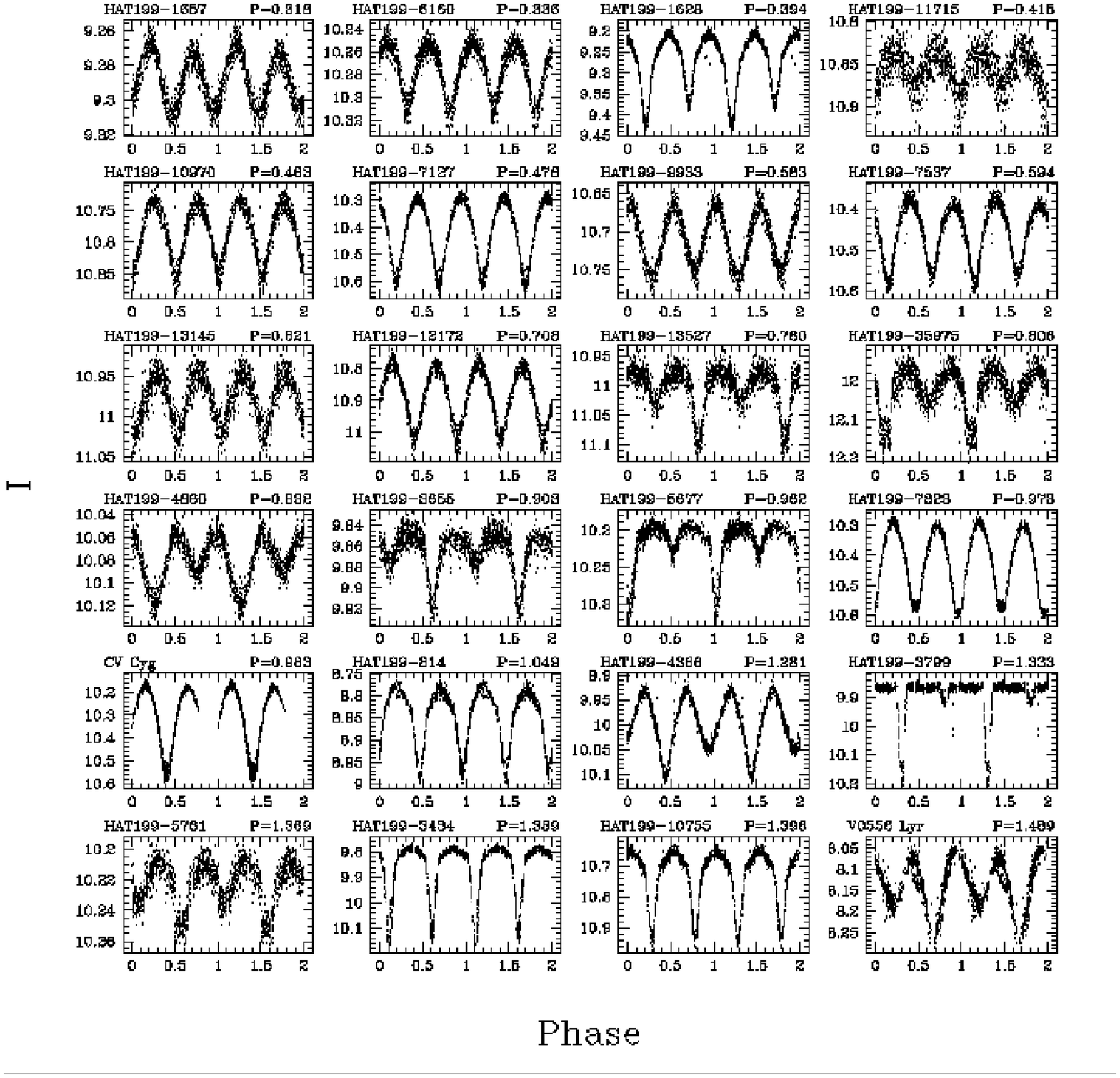}
\caption{I-band light curves, sorted by period (given in days), for 48 
of the 157 EBs in our catalog. {\it GCVS}\/ names are provided where
available.}
\label{EBfig}
\end{figure}

\addtocounter{figure}{-1}
\begin{figure}[p]
\epsscale{1.0}
\plotone{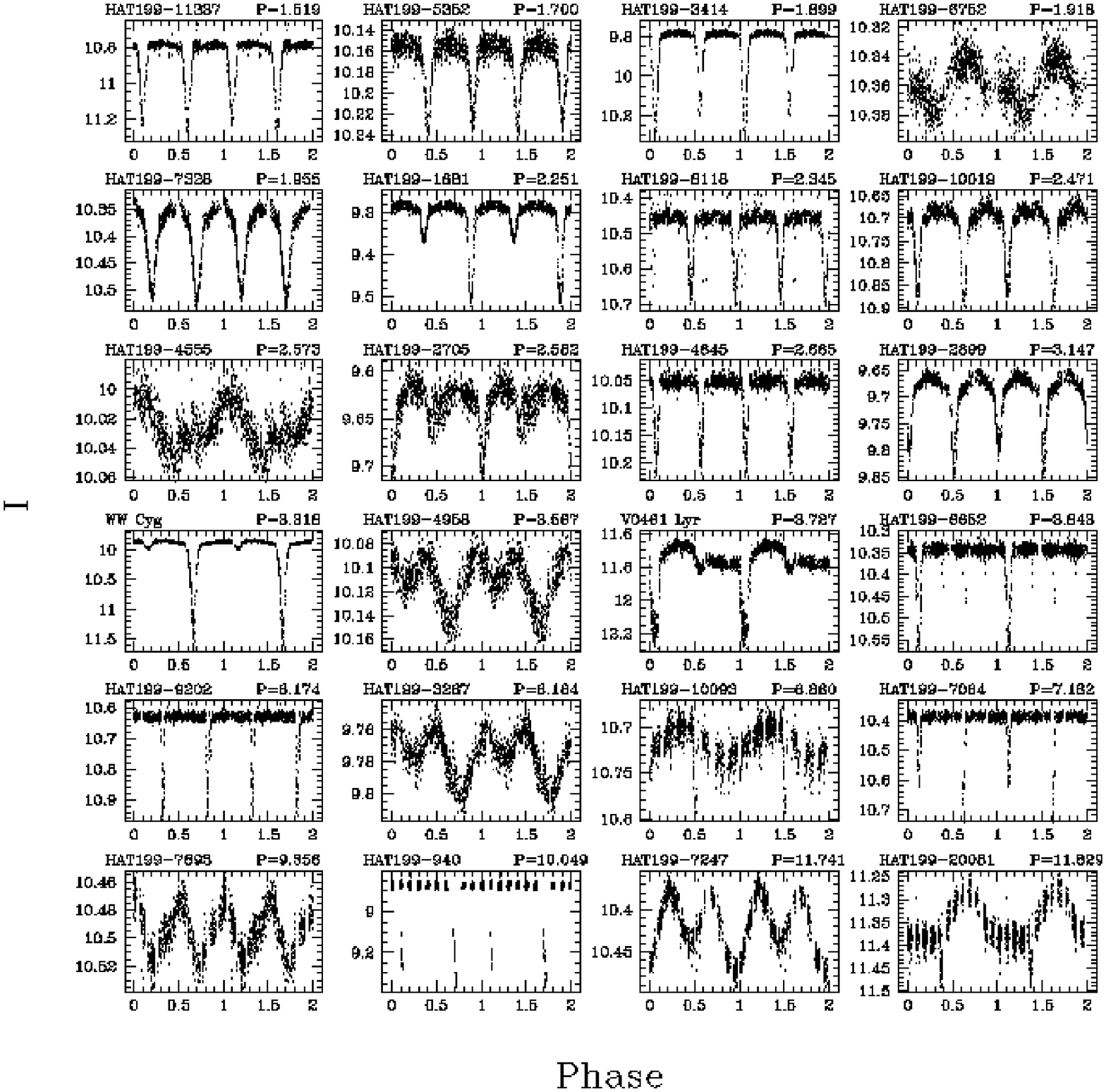}
\caption{{\it Continued}.}
\end{figure}

\begin{figure}[p]
\epsscale{1.0}
\plotone{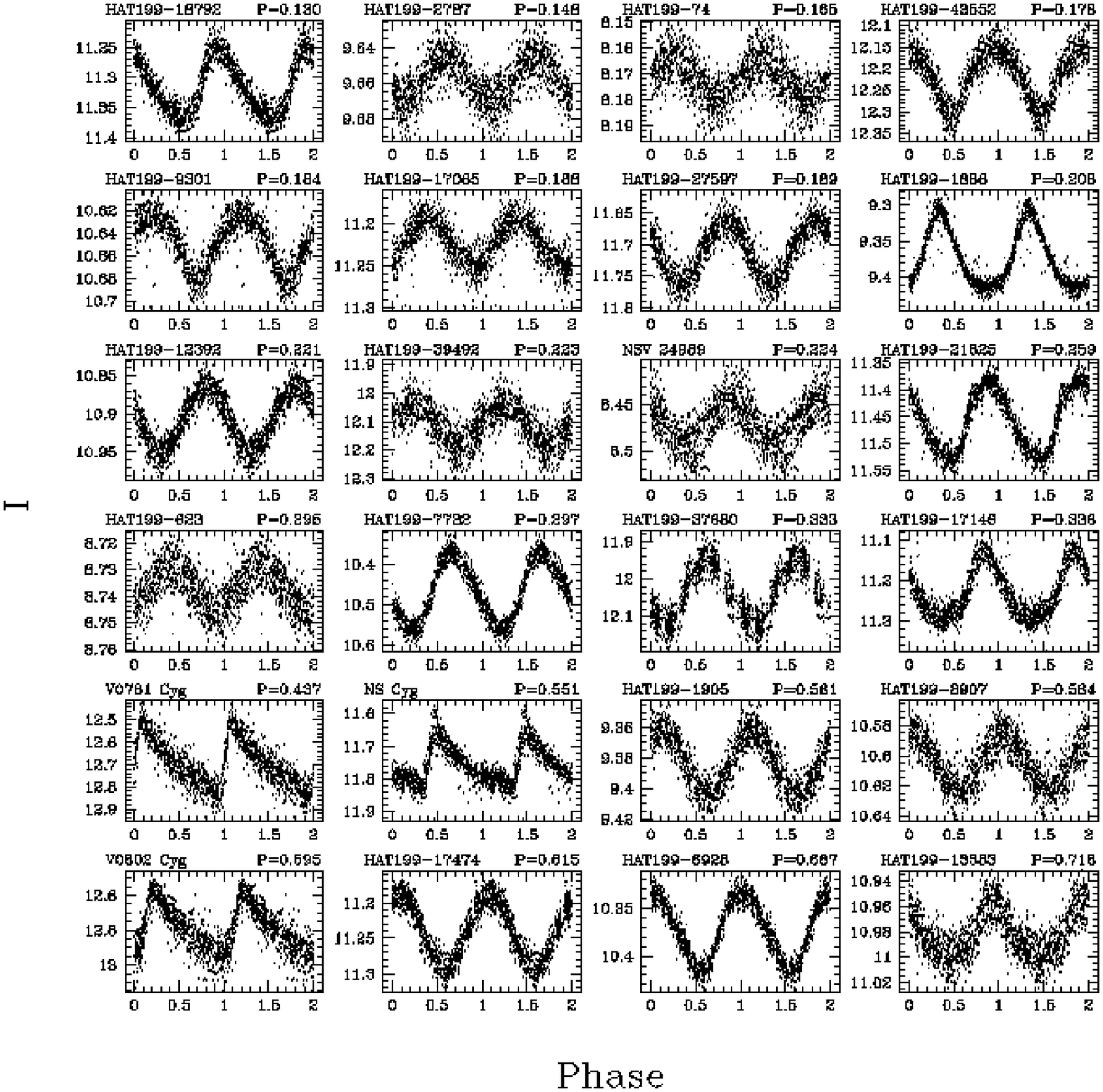}
\caption{I-band light curves, sorted by period (given in days), for 48 of 
the 50 pulsating variables in our catalog. The other two are shown in
Figure~\ref{multiperiod}.}
\label{PULfig}
\end{figure}

\addtocounter{figure}{-1}
\begin{figure}[p]
\epsscale{1.0}
\plotone{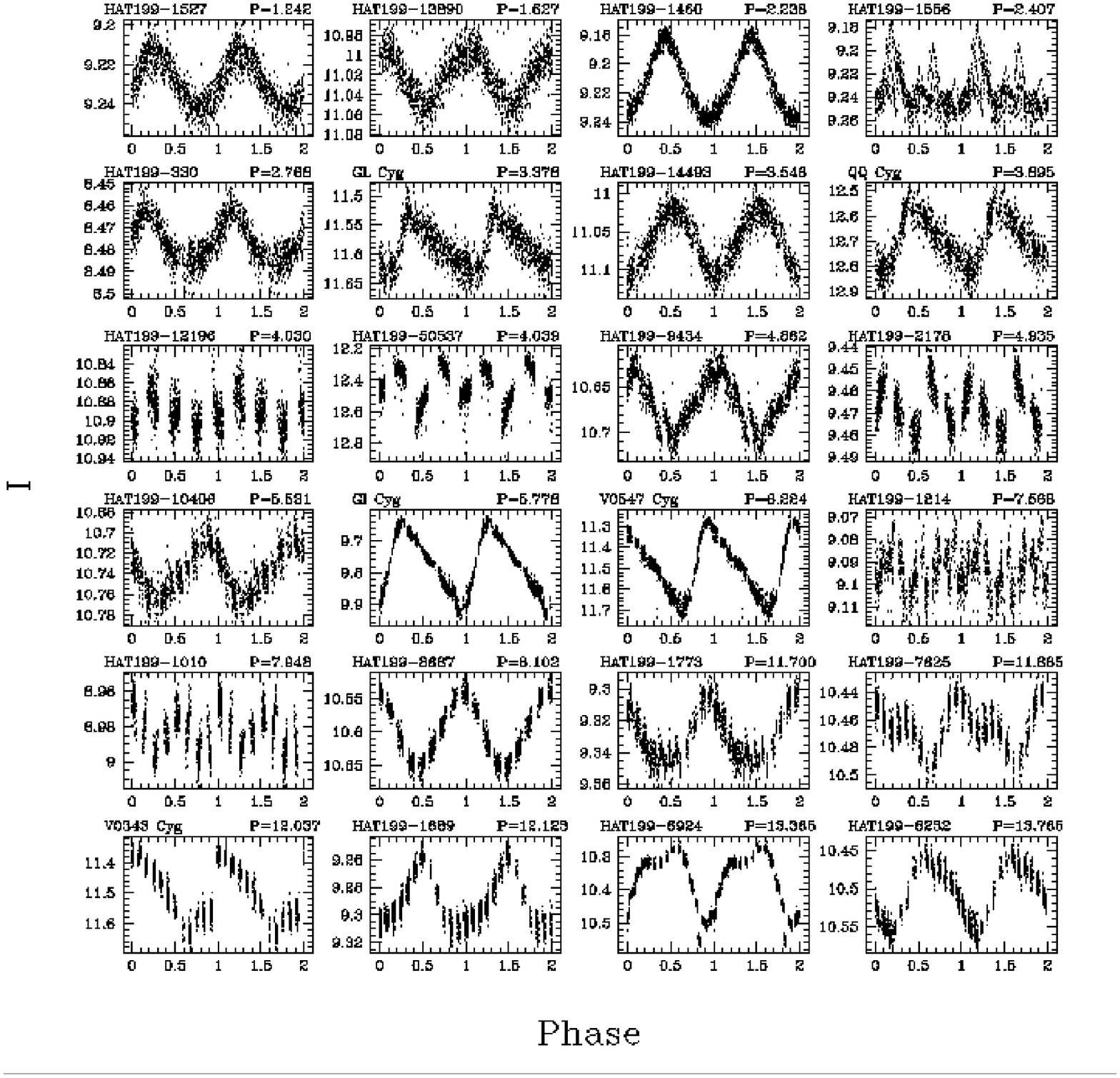}
\caption{{\it Continued}.}
\end{figure} 

Of the 30 periodic variables matched to {\it GCVS}\/ objects, 11 are
Algol type EBs, 3 are $\beta$ Lyr type EBs, 5 are WUMa type EBs, 3 are
Cepheids, 1 is a population II Cepheid, and 4 are RR Lyr
variables. One short period (0.224 days) pulsating variable has a
match in the {\it NSV} catalog. The other known periodic variable
which has no counterpart in the {\it GCVS}\/ is HD227269 (see \S5.1).

\begin{figure}[p]
\epsscale{1}
\plotone{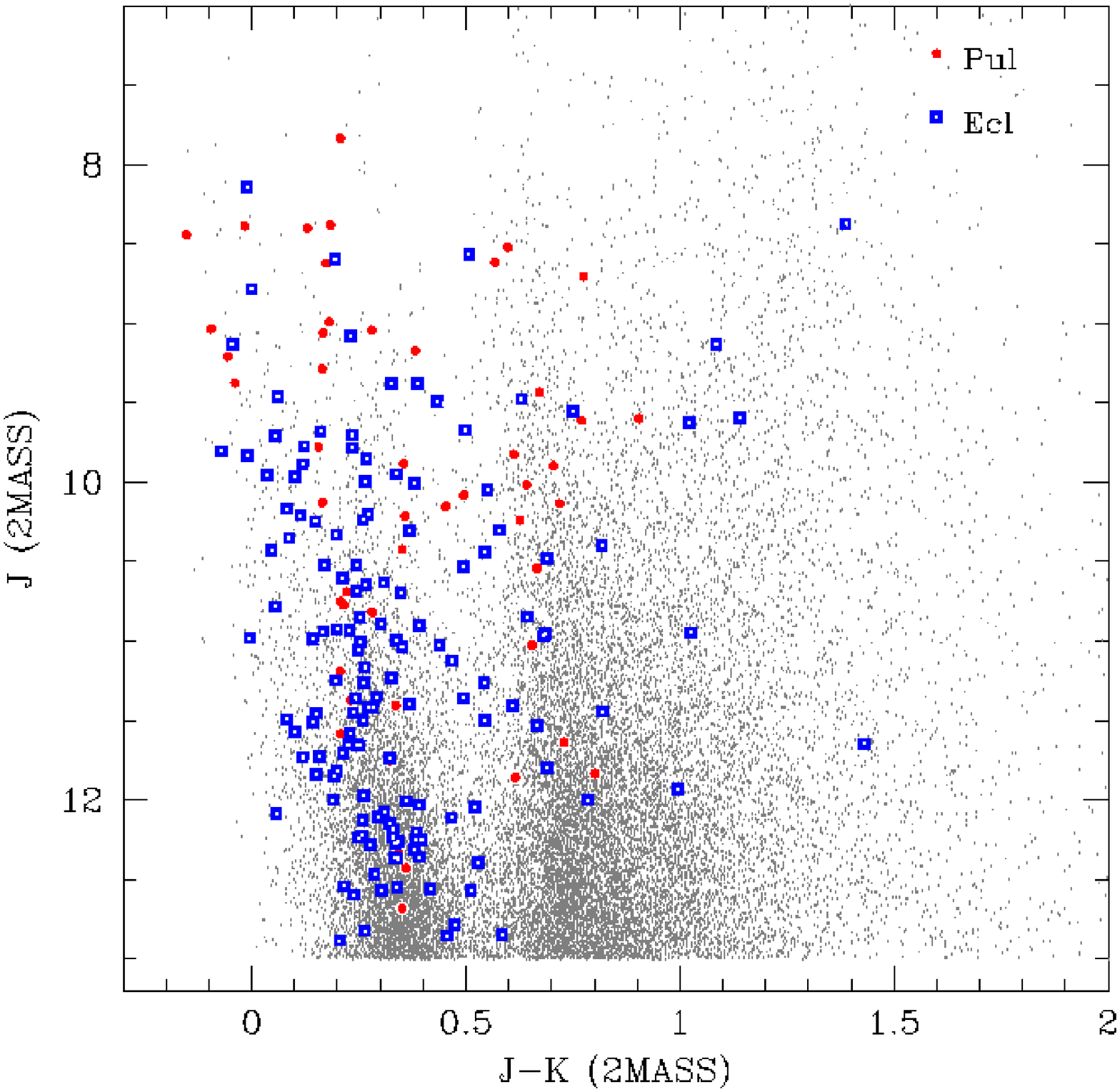}
\caption{J vs. J-K CMD showing location of EB and pulsating variables
relative to all the {\it 2MASS}\/ objects in our field with $J<13$.
Magnitudes are taken from {\it 2MASS}. Note that the axis ranges are
not the same as in Fig~\ref{cmd1}. Pulsating variables are shown as
dots, EBs as squares, and the general population of objects is shown in
grey. Note that both classes of periodic variables tend to lie toward
the blue end (on the main sequence) relative to the LPVs and
miscellaneous variables (Fig.~\ref{cmd1}).}
\label{cmd2}
\end{figure}

Figure~\ref{cmd2} shows the location of the periodic variables
(separated into pulsating and eclipsing categories) on a J vs.~J-K
CMD.  Compared to the LPVs and miscellaneous variables
(Fig~\ref{cmd1}) these objects tend to be blue. For the pulsating
stars this is expected as we are only classifying stars with periods
less than 15 days as periodic, and the shorter period stars tend to be
denser and hotter (e.g.~$\delta$ Scuti).

The two most studied of the matched objects are WW Cyg, an EA/SD
binary, and CV Cyg, a EW/DW binary. Because WW Cyg shows very deep
primary eclipses (3.5 mag in V) it has been frequently observed over
the last century. As a result this system has been particularly useful
in probing the period changes in close binaries (Zavala et al.\ 2002).
Our observations reveal much shallower eclipses in I (2.05 mag) and
clearly reveal the secondary eclipses (0.15 mag in I) which have
hitherto been undetected in V. Struve\ (1946) determined that the
primary has spectral type B8, while Yoon et al.\ (1994) assigned a
spectroscopic type G9 to the secondary. The difference of $\sim1.5$ in
the amplitude of the primary eclipse in V and I suggests that the
secondary should have a later spectral type (later than K3). However
Yoon et al.~point out that the photometrically determined spectral
types for Algol secondaries are typically later than the
spectroscopically determined ones.

CV Cyg is a highly evolved eclipsing system that has been used
primarily for the study of period and amplitude changes in close
binaries (Demircan et al.\ 1995). Because the period is nearly 1 day
(0.983 days) we do not observe both the primary and secondary eclipses,
however we do obtain approximately 15 minima observations.

\begin{figure}[p]
\epsscale{1}
\plotone{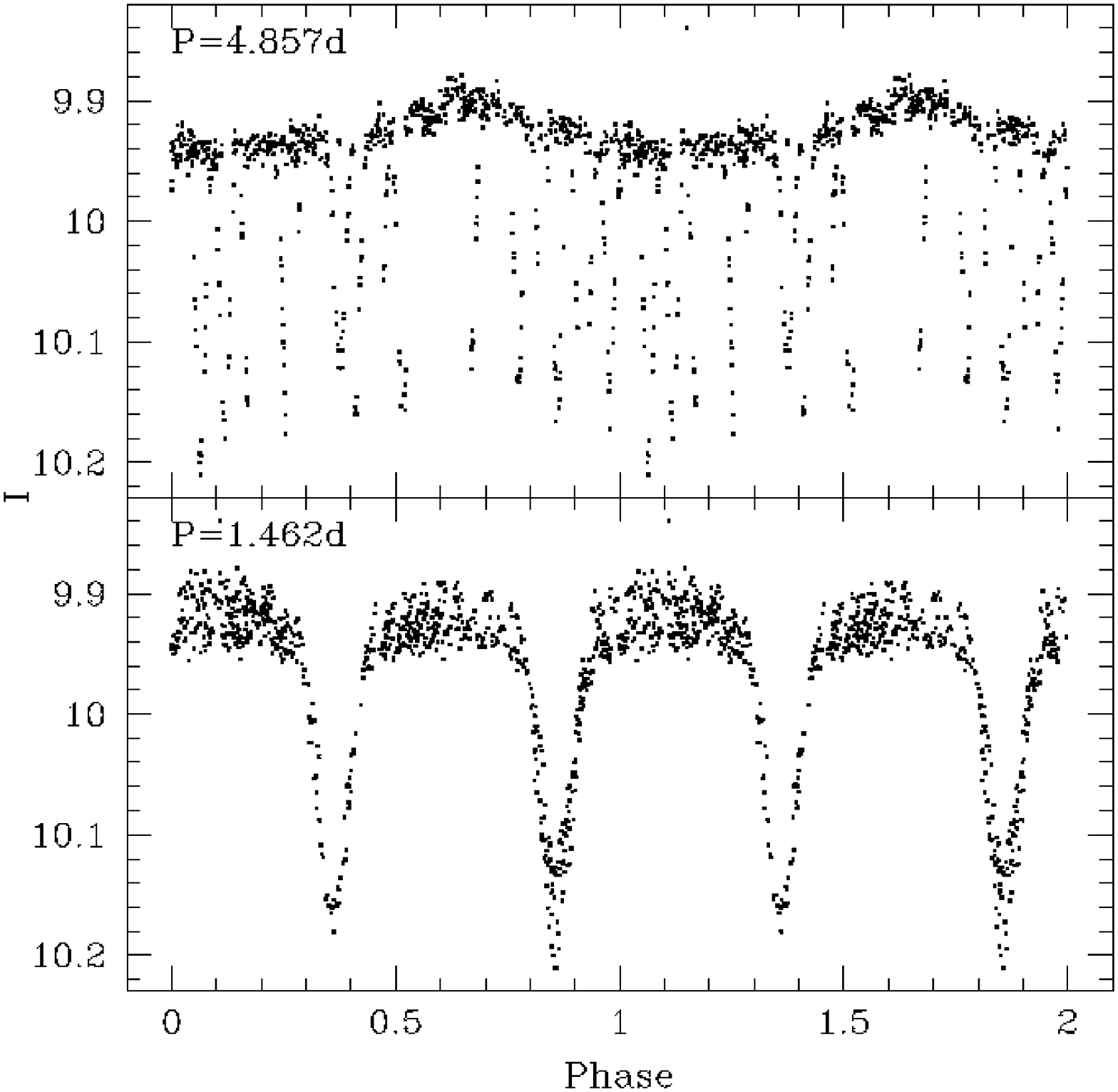}
\caption{Light curve of V1171 Cyg shows eclipses with an orbital
period of 1.462 days and a Cepheid-like upper envelope modulation with
a period of 4.857 days (See discussion in \S5.3).}
\label{lc4144_mag}
\end{figure}

The light curve of one very interesting object that we observe in shown
in Figure~\ref{lc4144_mag}. This object matches the known EA/KE binary
V1171 Cyg discovered by Wachmann (1966). It has also been detected as a
visual double with 0\farcs34 separation by Couteau\ (1981). The system
has spectral type B9 as listed on SIMBAD. We obtain an orbital period
for this system of P=1.462d, and also observe an upper envelope
modulation with period P=4.857d and full-amplitude
$\sim0.05$ mag in I. To further analyze this system we obtained 
spectroscopy using the FLWO 1.5m telescope. The spectra show color
changes, which indicate that this is not a random blend with another
variable more than $1\arcsec$ away. The light curve appears to be very
similar to HD227269 (\S5.1 and Fig.~\ref{lc2743_mag.data}), however in
this case we have greater confidence that the pulsations and eclipses
are occurring in the same system. We suspect that we are looking at a
triple system with a Cepheid as one component. Evans et al.\ (2003)
have found that a large fraction of Cepheids exist in triple systems,
but have been unable to determine the masses of all 3 stars in a given
system. If this system is indeed a triple, it may be possible to
measure all three masses and thereby add to our picture of the
distribution of masses among massive multiple systems (N.~Evans,
private communication).

We have also observed a number objects which appear to be short period
pulsating stars. These objects have periods between 0.1 and 0.3 days,
and may well correspond to $\delta$ Scuti type variables. A number
of these objects show multiple, non-harmonic periods. This includes
HAT199-539 which appears to have at least two periods, one at
0.1069d and another at 0.1198d, and HAT199-5178 with periods at
0.1203d and 0.1367d (Fig.~\ref{multiperiod}).

\begin{figure}[p]
\epsscale{1}
\plotone{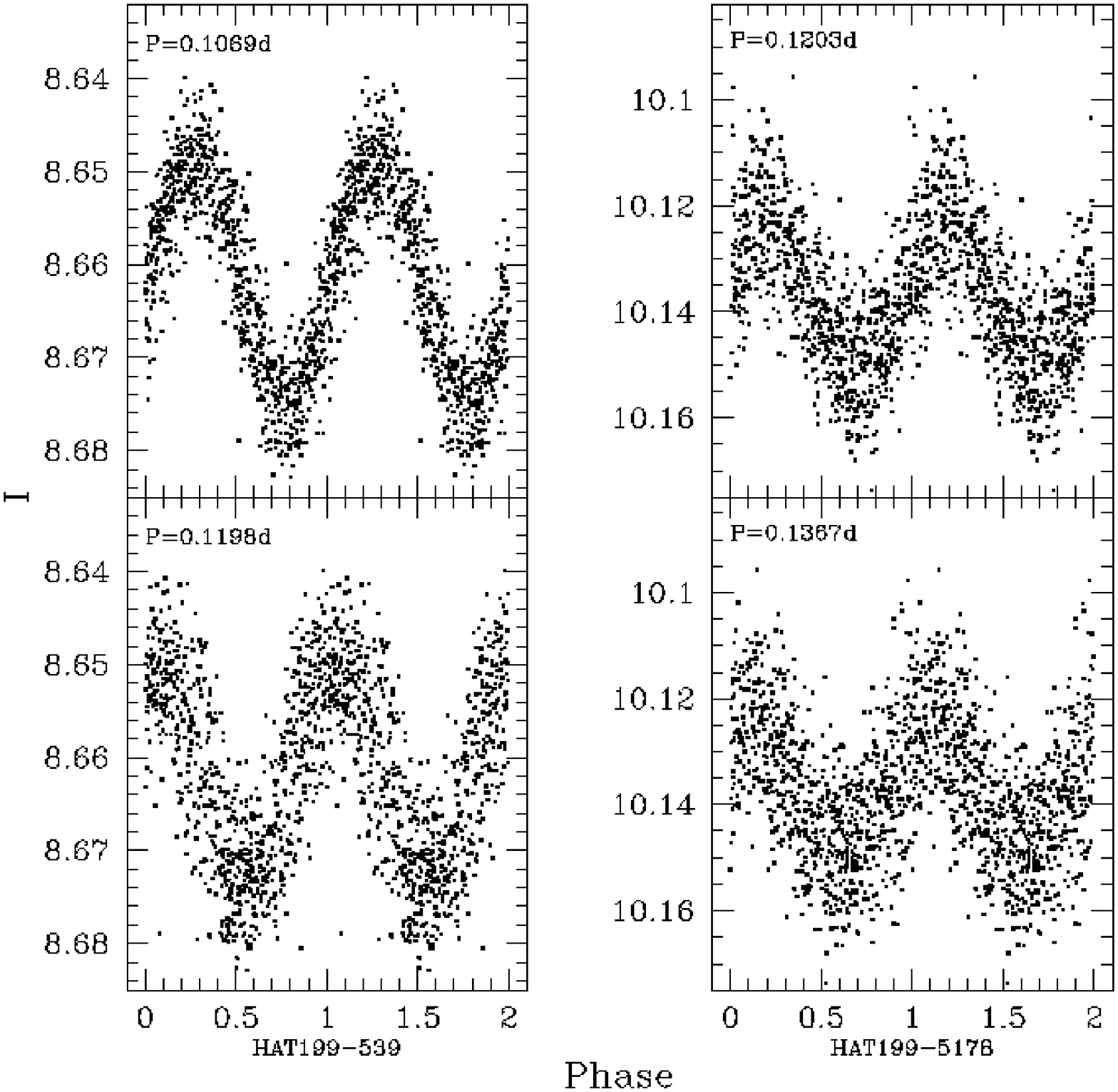}
\caption{Light curves of HAT199-539 and HAT199-5178, two short period 
pulsating variables that show multiple, non-harmonic periods. (Left)
HAT199-539 has periods of 0.1069d and 0.1198d. (Right) HAT199-5178 has
periods of 0.1203d and 0.1367d.}
\label{multiperiod}
\end{figure}

\subsection{Miscellaneous Variables}

\begin{figure}[p]
\epsscale{1.0}
\plotone{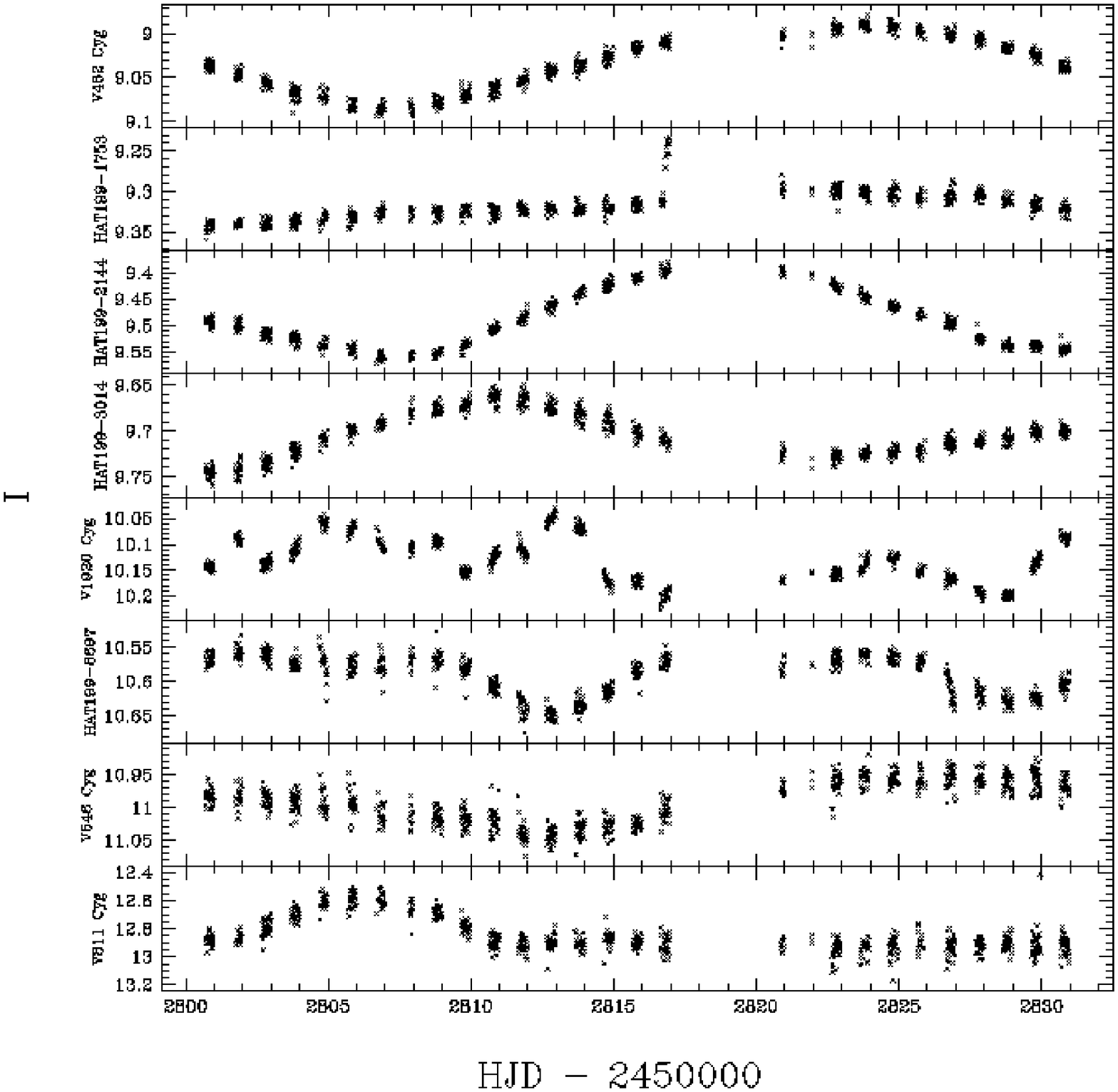}
\caption{I-band light curves for 8 of the 240 miscellaneous variables
in our catalog.}
\label{MISCfig}
\end{figure}

Variables which were not selected as LPVs and for which the best fit
period was longer than 14 days were classified as miscellaneous. This
classification may include a number of periodic variables with periods
typically between 14-30 days, as well as a number of irregular
variables that have timescales shorter than 30 days. We identified 241
such cases, of which all but 4 are newly discovered. The four matched
cases include: V482 Cyg, an RCB star, V1920 Cyg, a PV Telescopii
variable, V546 Cyg, an Algol-like eclipsing binary, and V811 Cyg, an SS
Cyg type dwarf nova. Light curves of these, and other interesting
miscellaneous variables are shown in Fig~\ref{MISCfig}.

V482 Cyg is a well studied member of the rare class of stars known as
R Coronae Borealis (RCB) variables. These carbon-rich variables are
noted for their unpredictable and substantial drops in brightness that
are attributed to the formation of soot clouds in the stellar
atmosphere. V482 Cyg was in the quiescent state during the
observations, however we do observe a pulsation-like light curve that
suggests a period of around 30 days (Fig~\ref{MISCfig}). This
oscillation is very similar to the 39-day period, Cepheid-like
pulsations observed in RY Sgr (see for example Lawson \& Cottrell,
1990).

V1920 Cyg is an Extreme Helium Star which Morrison \& Willingale\
(1987) observed to vary with an amplitude of 0.07 mag in V and a
period of 3 to 4 days. Fadeyev\ (1990) interpreted these variations as
pulsations in the second or higher overtone, and used them to
constrain the absolute magnitude of the star. Our observations of
V1920 Cyg over 30 days reveal irregular variations with no discernible
period, but with a timescale of roughly 4 days. We observe a maximum
full-amplitude of 0.2 mag in I (Fig~\ref{MISCfig}). This variable is
the bluest miscellaneous variable shown in Fig~\ref{cmd1}.

V811 Cyg is classified in the {\it GCVS} as a UGSS type variable. These
stars are dwarf novae that show regular, symmetric outbursts typically
lasting 3-10 days. We observe one such outburst for V811 Cyg with a
time-span of roughly 10 days and amplitude of $\sim0.35$ in I. It is
interesting to note that our observations of V811 Cyg have a quiescent
I of $\sim 12.90$ whereas Spogli et al.\ (2002) measured $I_{c} > 14.8$
in 1995. A likely interpretation is that we have observed a blend
between V811 Cyg and another brighter source. Although the light curve
we identify with V811 Cyg is one of the variables that does not have a
match with {\it 2MASS}, we do identify two sources in
the full {\it 2MASS}\/ catalog that lie within 10\arcsec~of our
coordinates for the object. These sources have $J=14.090$,
$J=15.397$, and $J-H>0$. A blend of these two objects in DAOPHOT/ALLSTAR
into a single object can account for the $\sim 12.90$ quiescent I. Note
that the coordinates for V811 Cyg in the {\it GCVS} are more than
10\arcsec~from our coordinates for the object.

One interesting miscellaneous variable that has not previously been
detected as a variable is HAT199-1753 which shows a flare-like
brightening by at least 0.08 mag over the course of 5
hours. Unfortunately the light curve is interrupted by several bad nights.

\subsection{Low Amplitude Periodic Variables}

\begin{figure}[p]
\epsscale{1}
\plotone{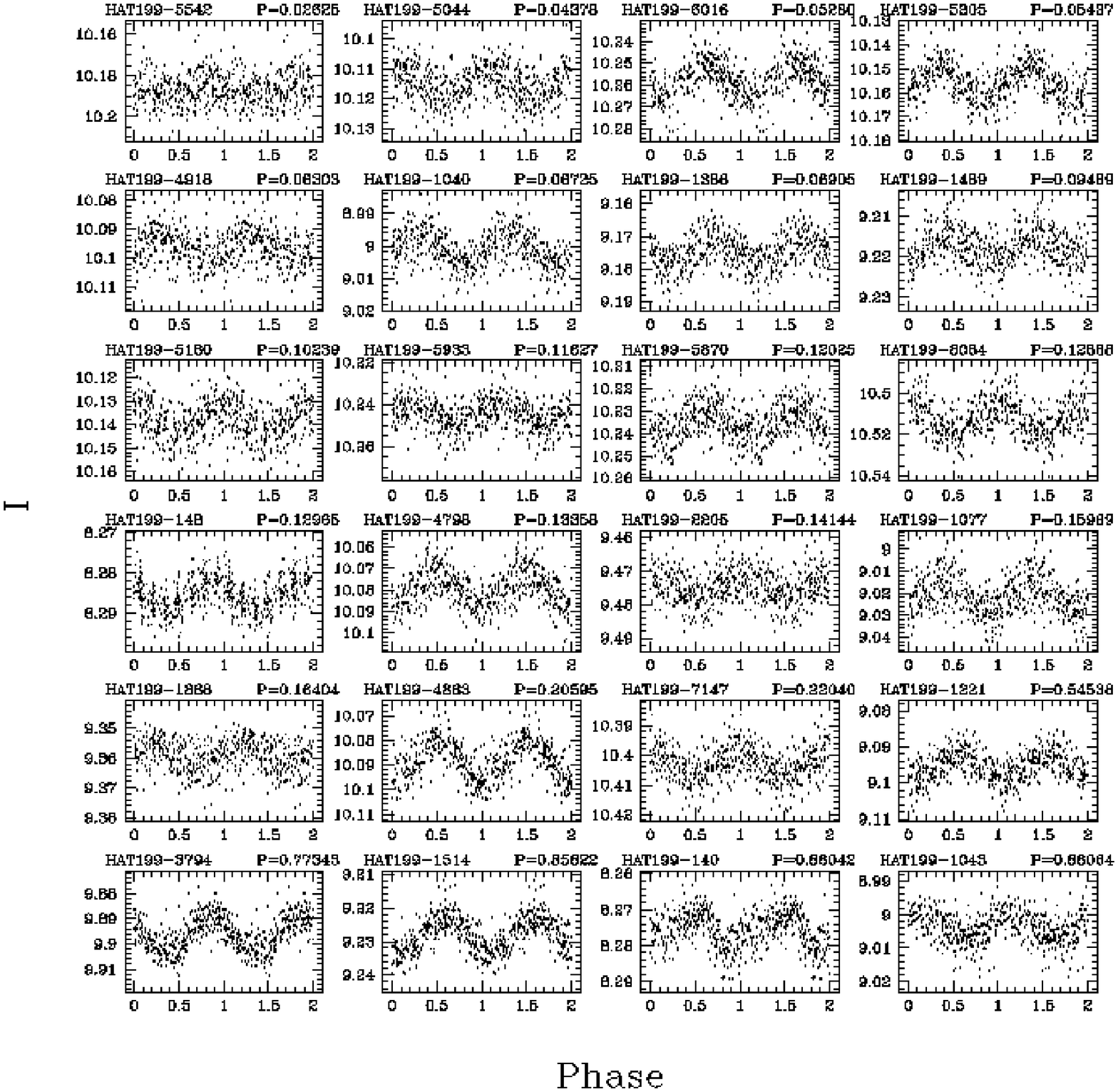}
\caption{I-band light curves of 24 low-amplitude, periodic variables
selected using the Schwarzenberg-Czerny algorithm.}
\label{FIGlowamp}
\end{figure}

As a test of our ability to detect very low amplitude periodic
variables we performed the Schwarzenberg-Czerny period finding test
(see \S 5.3) on the 8,949 light curves that had $RMS < 10$ mmag and
$J_{s} < 1.0$. Using this scheme we selected an additional 71 light
curves that had $P<0.9$ days, and $\sigma_{AoV}>3.5$ for the best
period, and did not lie near the 1/2 or 1/3 harmonic of one day. We
only selected objects that were more than 6 pixels from a variable in
our catalog to avoid issues of variability blending. We then analyzed
these light curves by eye and selected a list of 29 probable
low-amplitude pulsating variables. We show light curves for a few of
these in Figure~\ref{FIGlowamp}, some of which have full-amplitudes
approaching 10 mmag.

This is not a systematic search for these variables; we present the
results simply to demonstrate our capability of finding very low
amplitude variables in a high density field. We will continue
observations of this field and present results from a systematic search
for transits and other low amplitude variability in a future paper.

\section{Conclusion}

By using image subtraction photometry we have obtained light curves
for over 98,000 objects in a single field near the galactic
plane. From these light curves we have identified 1617 variable stars
with amplitudes greater than $\sim0.032$ mag, of which 1439 are
new. These include 1026 new LPVs, 176 new periodic variables and 237
new miscellaneous variables. The fact that 89\% of the variables were
previously undetected further demonstrates the vast number of
variables yet to be discovered even among fairly bright stars in our
Galaxy (Paczy\'{n}ski\ 1997).

We will continue our observations of this field in the hopes of detecting
planetary transits as well as other low amplitude variables by means
of extending our baseline. Using rudimentary selection techniques we
have already identified as many as 29 periodic variables with
amplitudes less than 0.05 mag. Besides the detection of low-amplitude
variables, future observations should also help in determining the
nature of the variables we have already discovered.

\acknowledgments{We gratefully acknowledge G.~Pojmanski for his
excellent ``lc'' program, W.~Pych for his ``fwhm'' program, J.~Devor
for his period-finding code, P.~Berlind, M.~Calkins, and T.~Matheson
for obtaining and analyzing spectra of V1171 Cyg, D.~Latham and
N.~Evans for helpful discussion, A.~Bonanos, B.~Mochejska and C.~Alard
for their help with the ISIS photometric package, and P.~Wils for his
help with matching to the GCVS and NSV catalogs. We are grateful to
C.~Akerlof and the ROTSE project for the long-term loan of the lens
used in the HAT-5 telescope. We also thank R.~Lupton for his help and
useful suggestions in refereeing this paper. This research has made
use of the SIMBAD database, operated at CDS, Strasbourg, France. JDH
is funded by a National Science Foundation Graduate Student Research
Fellowship. Partial support for the HAT project has been provided
through NASA grant NAG 5-10854.}

\appendix

\section{Running Image Subtraction}

To select a reference for alignment we measured the full-width at
half-maximum (FWHM) values of all the images and then chose an image
which had one of the smallest FWHMs, circular profiles, and had good
spatial overlap with the remaining images. At this point we also
examined images with extreme values for FWHM (typically frames for
which the program failed). Many of these turned out to be partially
obscured by clouds, or have irregular background patterns whose cause
could not always be determined. These images were removed before
proceeding with registration/subtraction.

Before proceeding with subtraction we had to establish a set of
subtraction parameters. These parameters include the number of
independent regions in which to divide the images ({\tt sub\_x} and
{\tt sub\_y}), the form of the kernel to use (including the number of
gaussians, the ``width'' of the gaussians, the orders of the
polynomials associated with each Gaussian, the order of the polynomial
used to fit spatial variations in the kernel, and the order of the
polynomial used to fit spatial variations in the background), and the
sizes (in pixels) of the regions used for fitting the kernel ({\tt
half\_stamp\_size}) as well as for performing the PSF transformation
with the kernel ({\tt half\_mesh\_size}). We first created a
preliminary reference image from 30 of the lowest FWHM images, using a
trial guess for the parameters. We then ran the ISIS routine
``subtract.csh'' using several permutations of the parameters (we
varied the number of subregions, and the degree of background and
spatial variations). For each permutation we calculated the standard
deviation and mean of the subtracted images. We then chose the set of
parameters which produced the lowest average standard deviation and
mean. These include internally dividing each frame into 25 independent
sub-frames ({\tt sub\_x = sub\_y = 5}), using a first order polynomial
to fit the background variations and first order polynomial to fit the
spatial variations in the kernel, fitting with 3 gaussians, that had
``widths'' of 0.7, 2.0 and 4.0 respectively, and associated
polynomials of order 6, 4 and 3 respectively, and choosing a {\tt
half\_mesh\_size} of 11 pixels and a {\tt half\_stamp\_size} of 19
pixels. The internal subdivision is the most important parameter in
our case. This is because the ISIS implementation of image subtraction
assumes flux conservation over the whole field from image to
image. For a wide FOV this does not generally hold (due to, for
example, differential atmospheric extinction over the field). It
becomes necessary to subdivide the field into regions over which the
flux conservation assumption holds within the limits set by photon
noise/sky background. We empirically find that subdividing into more
than 25 sections yields negligible improvement to the standard
deviation of the subtracted images (and the light curves of a selected
sample of stars) while significantly increasing the computation time.

After obtaining a final reference image (composed of 47 images) for
subtraction, we performed subtraction on all the images. At this point
we sorted the subtracted images by standard deviation and mean,
examining images with large standard deviations and/or means
significantly different from zero. This procedure allowed us to
identify images with subtle cloud patterns, etc., which might
contaminate the photometry. Since image subtraction assumes that the
flux of all stars scales by a constant from image to image, any complex
background variations/clouds will yield significant residuals over an
entire sub-region of the subtracted image. The ease with which very
subtle differences in images can be identified is one advantage of
using image subtraction over methods to directly measure photometry on
original images.

After cleaning, we were left with 800 out of 935 images on which to
proceed with photometry.

\section{Performing Photometry}

We obtained the light curves for 98,000 objects using the ``phot.csh''
routine contained in the ISIS package. This procedure first determines
a PSF within a region of size {\tt PSF\_width} pixels on the reference
image. This is done by taking the median profile of a stack of bright
stars. The profiles for each star are interpolated (using a cubic
B-spline) onto a common grid. Spatial variations in the PSF are
accounted for by splitting the image into several sub-areas. The
routine then uses the best-fit kernel to transform it into the PSF for
the subtracted image. It then measures the flux within a radius of
{\tt radphot} pixels from a specified location, weighting it by the
PSF, and normalizes by the integral of the PSF squared over a region
of radius {\tt rad\_aper} pixels. These difference flux measurements
can be converted into instrumental magnitude light curves for each
object using the flux on the reference image measured by
DAOPHOT/ALLSTAR.

In converting from flux to magnitudes it is important to ensure that
the flux measured by DAOPHOT/ALLSTAR is properly scaled to the flux
measured by ISIS. The DAOPHOT/ALLSTAR fluxes correspond to the flux
from PSF fitting using a zero-point of 25.0 magnitudes. In general,
this flux is {\it not}\/ equal to the total number of ADUs in the
image contributed by the star in question. For a large aperture
radius, the latter value is what is measured by ``phot.csh.'' To
determine the scaling between the two ``fluxes'' we perform an
aperture correction as follows: after fitting a PSF to 98,000 stars in
the field, we subtract a number of these stars from the reference
image leaving only a sparsely populated image. We then perform
aperture photometry on the remaining stars in this sparse image (using
a large aperture radius of 7 pixels). Comparing the aperture
magnitudes with the PSF magnitudes we find that the PSF magnitudes are
consistently brighter by 0.19 mag. We can correct for this difference
in scaling by using a zero-point of $25.0 - 0.19 = 24.81$ magnitudes
for the DAOPHOT/ALLSTAR reference magnitudes of our program stars.

To optimize the ISIS photometry parameters we ran the ``phot.csh''
procedure iteratively on a small subset of the 98,000 objects with
magnitudes across the entire range. The parameters we varied include
({\tt rad\_aper}), the {\tt PSF\_width}, and the inner/outer radii of
the annulus for measuring the background ({\tt rad1\_bg} and {\tt
rad2\_bg}). We related the parameters so that $PSF\_width =
2*rad\_aper+3$, $rad1\_bg = 2*rad\_aper+1$ and $rad2\_bg=
2*rad\_aper+6$. We also independently varied the radius for photometry
({\tt radphot}) with little effect. We find that when the aperture
radius is reduced below 7 pixels for our FWHM of $\sim$~3 pixels the
amplitude of the light curves (including the RMS) is artificially
reduced, and the light curves of variable stars (especially large
amplitude variable stars) become exessively noisy. We believe that
this is the result inaccurately determining the PSF (increasing the
noise of the variable stars) while simultaneously providing a flux
that is not properly scaled (yielding lower amplitudes). For {\tt
rad\_aper} greater than 7 pixels the effect on the light curves is
minimal, with a slight increase in the noise for the dim stars due to
sampling more of the background. The final set of photometry
parameters that we chose include: {\tt rad\_aper = 7 pixels,
PSF\_width = 17 pixels, radphot = 3.0 pixels, rad1\_bg = 15.0 pixels}
and {\tt rad2\_bg = 20.0 pixels}.

After obtaining light curves for all the stars in our field, we
checked to see if there were any remaining bad frames. To do this we
calculated the RMS of each light curve and then calculated, for each
frame, the number of light curves for which the magnitude of that
frame was more than 3-$\sigma$ from the mean. We identified 12 frames
that were consistently ``bad'' in a large number of light
curves. These frames were contaminated by clouds, or airplane tracks,
or simply had subtracted images with substantial residuals and
background gradients whose cause could not be identified on the
original image. After removing these frames from every light curve we
were left with a total of 788 contributing frames. Several of the
light curves contained an excessive number of outlier points as a
result of lying near the edge of the field. To clean these light
curves we removed any point with a formal flux error less than 15 ADU
as determined by the ``phot.csh'' routine (the average flux error
being $\sim$~40 ADU for the dimmest stars). Some light curves also
contained bad points as a result of lying near a saturated
star. Because the profile of saturated stars is not fit by the PSF,
these stars yield significant residuals on the subtracted
images. Since these residuals change from image to image the
contribution of this residual to the aperture of a nearby star
changes.  For dim stars this effect dominates such that occasionally
the difference flux measured by ``phot.csh'' is more negative than the
flux of the star itself as measured by DAOPHOT. These can be
eliminated by removing all points with negative infinity magnitude
from the light curves. The variability induced by proximity to
saturated stars is difficult to distinguish from actual variability,
we have attempted to account for these using templates to remove
``bad'' light curves (\S4.4).

\newpage

\begin{deluxetable}{lrrrrrrc}
\tabletypesize{\footnotesize}
\tablewidth{0pc}
\tablecaption{HAT Catalog of ``Kepler's Field'' Variables: first 8 columns.}
\tablehead{\colhead{ID [HAT199-]} & \colhead{$\alpha_{2000}$} &
\colhead{$\delta_{2000}$} &\colhead{I} & \colhead{J} &
\colhead{H} &\colhead{K} & \colhead{{\it 2MASS} ID}}
\startdata
00001 &  19$^{h}$44$^{m}$49\fs29 & 37\degr32\arcmin59\farcs6 &  07.787 &  05.601 &  04.760 &  04.251 &  1275.128417 \\
00006 &  19$^{h}$40$^{m}$59\fs04 & 36\degr43\arcmin32\farcs8 &  07.959 &  04.039 &  02.999 &  02.461 &  1267.127829 \\
00009 &  19$^{h}$46$^{m}$42\fs35 & 34\degr50\arcmin40\farcs6 &  07.989 &  05.097 &  04.212 &  03.875 &  1267.127829 \\
00020 &  20$^{h}$03$^{m}$57\fs48 & 39\degr59\arcmin16\farcs7 &  08.063 &  05.494 &  04.491 &  04.093 &  1299.142572 \\
00029 &  20$^{h}$01$^{m}$50\fs00 & 33\degr28\arcmin24\farcs0 &  08.090 &  \nodata &  \nodata &  \nodata &  \nodata \\ 
00049 &  19$^{h}$43$^{m}$09\fs81 & 34\degr06\arcmin09\farcs6 &  08.137 &  06.050 &  05.146 &  04.785 &  1241.138907 \\ 
00060 &  19$^{h}$25$^{m}$08\fs27 & 35\degr59\arcmin57\farcs9 &  08.151 &  08.142 &  08.171 &  08.154 &  1259.119674 \\
00061 &  19$^{h}$41$^{m}$17\fs60 & 40\degr10\arcmin41\farcs9 &  08.154 &  06.758 &  05.819 &  05.538 &  1301.125923 \\ 
00074 &  19$^{h}$49$^{m}$59\fs62 & 35\degr40\arcmin14\farcs5 &  08.173 &  07.833 &  07.700 &  07.627 &  1256.139193 \\
\enddata
\tablecomments{ Coordinates are from {\it 2MASS} where available, as are J, H, and K measurements. The completed version of this table is in the electronic edition of the refereed paper.}
\label{tab1}
\end{deluxetable}

\begin{deluxetable}{lrrrrcc}
\tabletypesize{\footnotesize}
\tablewidth{0pc}
\tablecaption{HAT Catalog of ``Kepler's Field'' Variables: first column and last 6 columns.}
\tablehead{\colhead{ID [HAT199-]} & \colhead{CLASS} &
\colhead{$I_{min}$} &\colhead{$I_{max}$} & \colhead{P [days]} &
\colhead{{\it GCVS} ID} &\colhead{{\it GCVS} CLASS}}
\startdata
00001 &   LPV &  07.987 &  07.642 & \nodata & \nodata & \nodata \\
00006 &   LPV &  08.284 &  07.800 & \nodata & V942 Cyg & M \\
00009 &   LPV &  08.024 &  07.917 & \nodata & \nodata & \nodata \\
00020 &   LPV &  08.116 &  08.003 & \nodata & V423 Cyg & SRA \\
00029 &   LPV &  08.172 &  08.052 & \nodata & \nodata & \nodata \\
00049 &   LPV &  08.294 &  08.028 & \nodata & \nodata & \nodata \\
00060 &   ECL &  08.275 &  08.044 & 01.4891 & V556 Lyr & \nodata \\
00061 &   MIS &  08.179 &  08.124 & \nodata & \nodata & \nodata \\
00074 &   PUL &  08.192 &  08.153 & 00.1650 & \nodata & \nodata \\
\enddata
\tablecomments{Maximum and minimum I are the 6th from the brightest and dimmest measurements respectively. The completed version of this table is in the electronic edition of the refereed paper.}
\label{tab2}
\end{deluxetable}

\end{document}